\documentclass[showpacs,pre,numerical,10pt,superscriptaddress,notitlepage, twocolumn]{revtex4-1}
\usepackage{graphicx}
\usepackage{amsmath,amssymb}
\usepackage{times,mathptmx}
\usepackage{units}
\usepackage{epsfig}
\usepackage{booktabs}
\usepackage{multirow}

\usepackage[latin1]{inputenc}
\usepackage[english]{babel}
\usepackage{color}

\newcommand{\ii}{\mathrm{i}}
\renewcommand{\Re}{\operatorname{Re}}
\renewcommand{\Im}{\operatorname{Im}}

\definecolor{darkgreen}{rgb}{0,0.5,0}
  
\begin{document}

\date{\today}
\title{Bistability in Two Simple Symmetrically Coupled Oscillators \\with Symmetry-broken Amplitude- and Phase-Locking}

\author{Andr\'e R\"ohm}\affiliation{Institut f{\"u}r Theoretische Physik, Technische Universit{\"a}t Berlin, 10623 Berlin, Germany}
\author{Kathy L\"udge}\affiliation{Institut f{\"u}r Theoretische Physik, Technische Universit{\"a}t Berlin, 10623 Berlin, Germany}
\author{Isabelle Schneider}\affiliation{Institut f{\"u}r Mathematik, Freie Universit{\"a}t Berlin, 14195 Berlin, Germany}

\begin{abstract}
In the model system of two instantaneously and symmetrically coupled identical Stuart-Landau oscillators we demonstrate that there exist stable solutions with symmetry-broken amplitude- and phase-locking. 
These states are characterized by a non-trivial fixed phase or amplitude relationship between both oscillators, while simultaneously maintaining perfectly harmonic oscillations of the same frequency. 
While some of the surrounding bifurcations have been previously described, we present the first detailed analytical and numerical description of these states and present analytically and numerically how they are embedded in the bifurcation structure of the system, arising both from the in-phase and the anti-phase solutions, as well as through a saddle-node bifurcation. The dependence of both the amplitude and the phase on parameters can be expressed explicitly with analytic formulas. As opposed to previous reports, we find that these symmetry-broken states are stable, which can even be shown analytically.
As an example of symmetry-breaking solutions in a simple and symmetric system, these states have potential applications as bistable states for switches in a wide array of coupled oscillatory systems. 
\end{abstract}

\maketitle

\textbf{
The advantage of generic models in the nonlinear sciences is that they can approximate a wide variety of systems. Coupled oscillators are an important example. Often coupled oscillators possess an intrinsic symmetry, e.g. all \textsc{}oscillators being identical. In this case, symmetry-breaking bifurcations (sometimes also called 'spontaneous symmetry breaking') can create collective dynamics that no longer reflect the underlying structure of the system. In this paper we investigate and analytically describe a type of symmetry-breaking solution in a minimal oscillatory setup. Using the well-known Stuart-Landau oscillator, a model that can approximate a wide variety of real-world systems, we find symmetry-breaking solutions exhibiting harmonic oscillations. These patterns are naturally bistable, making them excellent candidates for use as a bistable switch. 
}

\section{Introduction}
\label{intro}

The harmonic oscillator is a universal model for systems with small oscillations.  Such systems are abundant in nature with different appearances. While the position and momentum oscillate in the classical mechanical pendulum, in LC circuits a periodic change between current and voltage is induced, and plain electromagnetic waves are made up of magnetic and electric fields reciprocally driving themselves.
However, many systems in nature exhibit a self-regulating amplitude which cannot be described by a linear harmonic oscillator. Examples include the population dynamics in predator-prey systems \cite{LOT20, VOL26a}, the periodic changes in the density of the chemical species in chemical oscillations \cite{BEL59, ZHA64}, or the electric field inside an active laser cavity \cite{ERN10b}. 
In all these cases, the oscillations converge towards a well-defined amplitude sustained by nonlinear interactions, i.e. a stable limit cycle exists.
The Stuart-Landau equation is a nonlinear extension of the harmonic oscillator and it can be thought of as the simplest possible nonlinear extension \cite{GAR12b} to describe these amplitude dynamics. Therefore the Stuart-Landau equation is arguably the most important model system for nonlinear oscillators.

The field of nonlinear dynamics is devoted to understanding the universal behavior of classes of systems. 
This is done by studying the topology of the state space for different parameters and connecting it with knowledge of the underlying bifurcations thereof. 
One of the most common and fundamental bifurcations is the Andronov-Hopf-bifurcation \cite{HOP42, AND49}, which describes the smooth onset of oscillations with a finite frequency. 
The Stuart-Landau oscillator is the normal-form of such a bifurcation \cite{KUR84, KUZ95}. 
Therefore, any nonlinear system close enough to an Andronov-Hopf-bifurcation can be approximated by a Stuart-Landau oscillator. 

While a single Stuart-Landau system is well understood, even such a simple system can lead to a rich zoo of dynamics in spatially extended versions. 
This can be done in two ways: 
First, if coupling is strong and oscillators can be thought of as a homogeneous field one obtains a system of PDE's, i.e., the Ginzburg-Landau-equations \cite{GAR12b}. 
Alternatively, when only a finite number of oscillators is considered, one arrives in the field of network or graph theory. 
We will be using the second approach in this paper. Real-world examples of networks of oscillators still cover a wide range of applications, numerical models and experiments from coupled lasers \cite{SOR13, GAR99a, JAV03, YAN04c, CLE14, KOZ00, ERZ05, ERZ09, HOH99b} to nanomechanical systems \cite{SHI07a}, chemical oscillators \cite{TOT15, WEI92}, coupled tunnel diodes \cite{HEI10} and the famous 17th~century study of synchronized mechanical clocks by Christiaan Huygens and modern versions thereof \cite{KAP12}. While many fascinating phenomena of coupled oscillators can be studied in pure phase-oscillators \cite{KUR02a, ABR04, DHU08} or in chaotic maps \cite{KAN90}, we believe the results of this paper to be mostly applicable to oscillators possessing more than a single degree of freedom.

The study of coupled Stuart-Landau oscillators has been a very active topic in the physics community for at least a decade. 
Among others, coupled Stuart-Landau oscillators have been shown to exhibit chaotic motion \cite{HAK92, NAK93, NAK94a, KU15}, amplitude and oscillation death \cite{KOS13, ZAK13}, coherence resonance \cite{USH05, GEF14}, chimera states \cite{ZAK14}, chimera death \cite{ZAK15b}, cluster states \cite{LEE13, LEH14, KU15} and more. Among bifurcation theorists in the mathematical community, the bifurcation structure of coupled Stuart-Landau oscillators has been thoroughly studied by Golubitsky and co-workers \cite{GOL88a}. Of note in this context is that the intrinsic connection between system solutions, bifurcations and the natural symmetries of the system is a valuable field of inquiry in itself. Especially the mathematical approach of equivariant bifurcation theory is a powerful tool for predicting likely dynamics of physical systems with inherent symmetries. However, even in the landmark case for this approach, i.e. coupled Stuart-Landau oscillators, one can still investigate novel patterns, which extend the established analytical and theoretical descriptions.
 
In this paper, we present new numerical results and analytical descriptions for a type of symmetry-breaking solution. These solutions exhibit a nontrivial symmetry-broken amplitude- and phase-locking while perfectly maintaining  rotational invariance. This is despite the fact, that the coupling we employ does not break any symmetries, and thus is not directly related to the usual Amplitude or Oscillation Death bifurcation scenarios \cite{KOS13}. These solutions have not been described in detail before, and we will therefore highlight their connection with the established bifurcation structures of coupled oscillatory systems. They are especially applicable to the case of lasers, where similar results have been obtained recently \cite{SEI17, CLE14}.

\begin{figure}
        \includegraphics[width=1.00\linewidth]{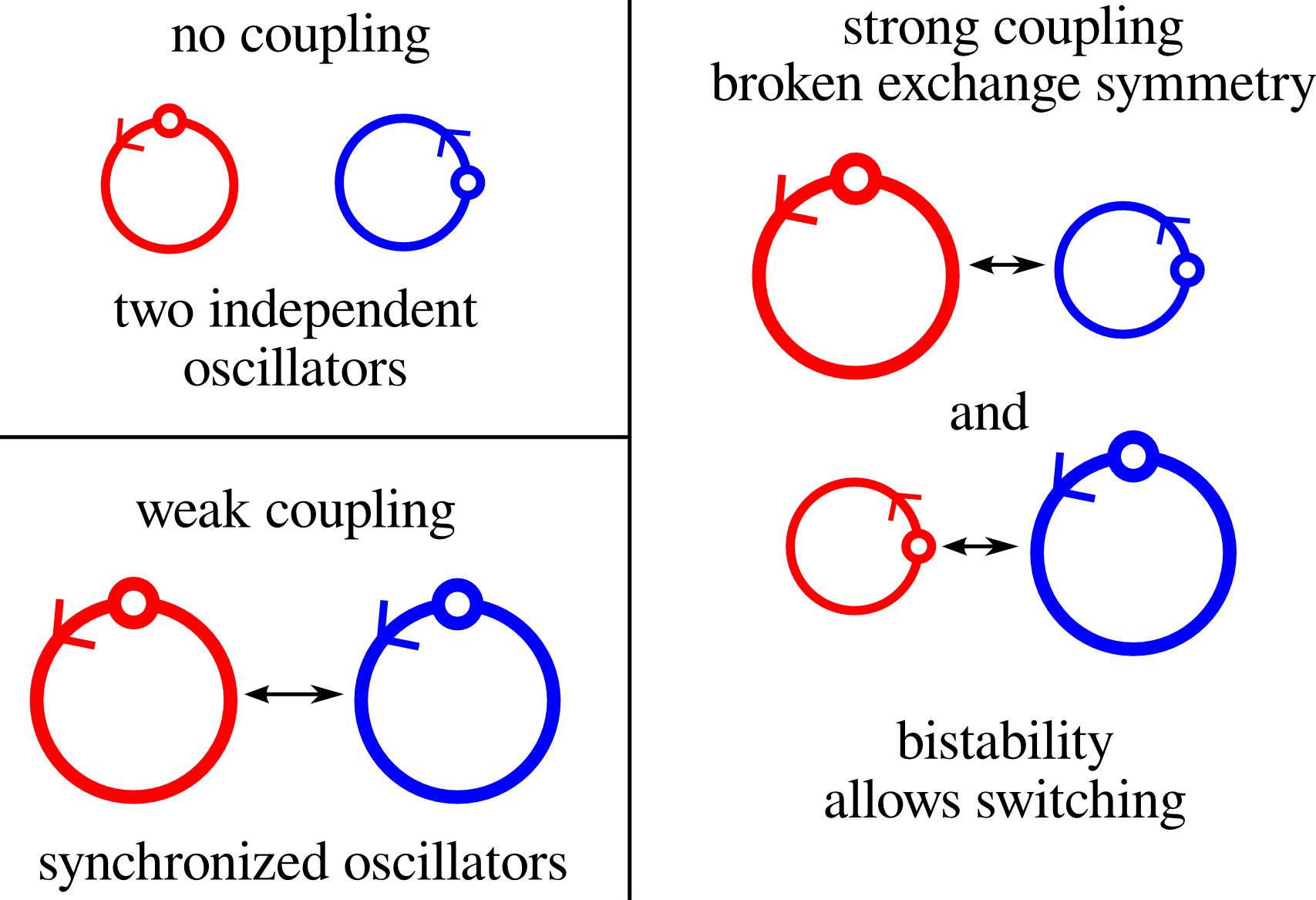}
        \caption{Sketch of the fundamental principles: When coupling two identical oscillators synchronization is easily achieved, even for low coupling strength. Increasing the coupling together with nonlinearities (i.e. shear) can lead to a symmetry-breaking. This naturally leads to bistability, which can be used as a switch.}
        \label{bistability_sketch}
\end{figure}

One of the main applications of bifurcation theory has always been the description of general bifurcation scenarios that can later be reproduced in more detailed models and experimental setups. One application that is obvious for the states found in this paper will be a bistable switch. For this, the Stuart-Landau system is especially suited because of it's generality. Fig.~\ref{bistability_sketch} illustrates the mechanism through which bistability can be generated by symmetry-breaking: While two identical oscillators are often found to synchronize when coupled symmetrically, this symmetry can be broken via subsequent bifurcations. When such a symmetry-broken state is achieved through sufficiently strong coupling, multistability follows naturally from the symmetries of the underlying system. This bistability can then be used as a switch. One can switch between both stable states with the help of a control signal, e.g. a pulse or a controlled disruption of the motion. An example of this basic principle can also be found in Ref.~\cite{CLE14}. 

\section{Model}

While it is known that time-delayed coupling and time-delayed self-feedback, as well as parameter mismatches and large or complex topologies can lead to a wide range of different dynamics, these extensions have the disadvantage of possessing a large number of parameters on which the dynamics critically depends. Furthermore, the more degrees of freedom are introduced into a system, the more can one expect to find a large number of solutions. 
In this paper, we want to address as many nonlinear systems as possible. We therefore focus on the fundamental case of two instantaneously coupled Stuart-Landau oscillators, described by the variables $Z_{1,2} \in \mathbb{C}$ in the following system of ODEs: 
\begin{align}        
    \dot{Z}_1 &=(\lambda + i\omega + \gamma\, |Z_1|^2) Z_1 + \kappa e^{i \phi} (Z_2 - Z_1)    
    \label{two_stuart_landau_equation_1}
    \\
    \dot{Z}_2&=(\lambda + i\omega + \gamma\, |Z_2|^2) Z_2 + \kappa e^{i \phi} (Z_1 - Z_2)  
    \label{two_stuart_landau_equation_2}
\end{align}
Here $\lambda \in \mathbb{R}$ is the bifurcation parameter with an Andronov-Hopf-bifurcation occurring at $\lambda = 0$ in a solitary oscillator, $\omega \in \mathbb{R}$ is the frequency of the free-running oscillator. 
The sign of the real part of the nonlinearity $\gamma \in \mathbb{C}$ defines whether the Andronov-Hopf-bifurcation is sub- or supercritical, while the imaginary part defines the hardness of the spring and induces an amplitude-phase coupling.
$\textbf{Im}(\gamma)$ is also linked to the amplitude-phase or linewidth-enhancement factor of semiconductor lasers \cite{BOE15}. 
We can assume that many nonlinear oscillators will have an effective $\textbf{Im}(\gamma) \not= 0$ since higher-order nonlinearities may lead to a similar coupling between phase and amplitude.
The coupling between the oscillators is defined by the coupling strength $\kappa \in \mathbb{R}$ and coupling phase $\phi \in [0, 2\pi]$. Later on we will also use the complex notation $\sigma = \kappa \exp (\ii \phi)$. For our numerical simulations we set $\mathbf{Re}(\gamma) = -0.1$ (supercritical case), $\mathbf{Im}(\gamma) = 0.5$, $\omega = 1$  and $\kappa = 0.1$, unless noted otherwise. 
This model can approximate a wide range of different oscillatory systems that are coupled instantaneously, i.e., with negligible transmission and coupling delay. 
Owing to the fact that the Stuart-Landau system is the normal form of an Andronov-Hopf-bifurcation, any system close to such a bifurcation can be approximated with the nonlinearity of Eqs.~\eqref{two_stuart_landau_equation_1} and~\eqref{two_stuart_landau_equation_2}. 
Note that by redefining $\lambda' = \lambda - \kappa \cos{\phi}$ and $\omega' = \omega  - \kappa \sin{\phi}$ one can also transform the coupling such that it does not include any self-feedback. Additionally, one can also either set $\kappa = 1$ or $\lambda = 1$  by a normalization of $Z$.


\section{Bifurcations and stable solutions}
\label{sec_bif}

\begin{figure*}[hbt]
        \includegraphics[width=1.00\linewidth]{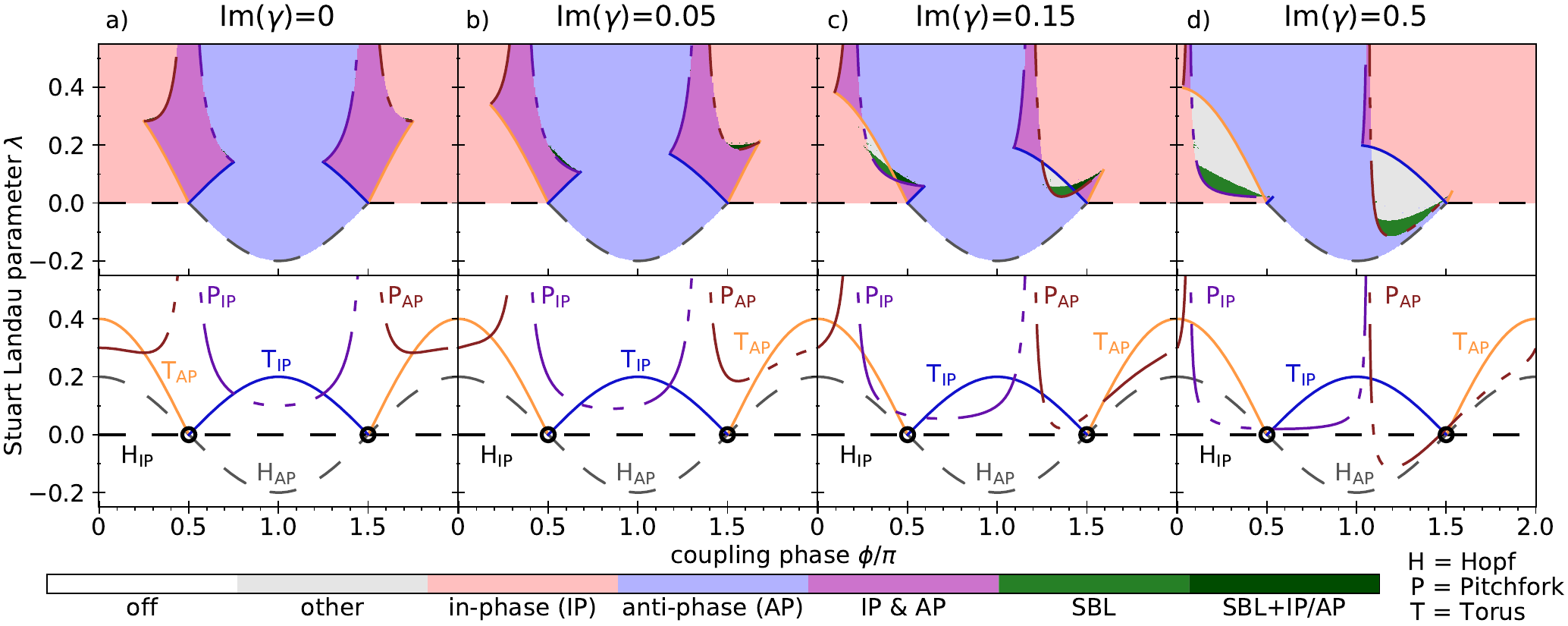}
        \caption{Numerically found states of two coupled Stuart-Landau oscillators as described in Eqs.~\eqref{two_stuart_landau_equation_1} and~\eqref{two_stuart_landau_equation_2} as a function of $\lambda$ and coupling phase $\phi$ for different $\mathbf{Im}(\gamma)$ . White areas correspond to the trivial off-state, light peach areas to the in-phase and bright blue areas to the anti-phase state, with their overlap in purple. Green corresponds to symmetry-broken amplitude- and phase-locking (SBL), while dark green marks this state with multistability either with the in-phase or anti-phase orbits (SBL+IP/AP). Light grey indicates higher order dynamics, e.g. quasiperiodic behavior. Lower row shows analytically derived bifurcation lines given by Eqs.~\eqref{HIP} and~\eqref{HAP} for the Hopf of the in-phase ($H_{IP}$) and anti-phase solution ($H_{AP}$), Eqs.~\eqref{TIP} and~\eqref{TAP} for the torus ($T_{IP}$ and $T_{AP}$), and Eqs.~\eqref{PIP} and~\eqref{PAP} for the Pitchfork bifurcation ($P_{IP}$ and $P_{AP}$). Black circles mark a co-dimension 2 bifurcation point. Parameters: $\mathbf{Re}(\gamma) = -0.1$, $\mathbf{Im}(\gamma)$ as indicated atop the columns, $\omega = 1$, $\kappa = 0.1$.}
        \label{2d_plot_multi}
\end{figure*}

From a mathematical viewpoint, Eqs.~\eqref{two_stuart_landau_equation_1} and~\eqref{two_stuart_landau_equation_2} are $\mathbb{Z}_2 \times S^1$-equivariant, where $\mathbb{Z}_2$ is
the cyclic group symmetry induced by the coupling between the single oscillators. The reflection is given by $Z_1 \leftrightarrow Z_2$ and the identity. $S^1$
is the rotational symmetry of the truncated Hopf normal form, i.e., $e^{i\theta}f(Z) = f(e^{i \theta} Z)$ for all angles $\theta \in  [0, 2\pi]$, where $f(Z)=(\lambda + i\omega + \gamma\, |Z|^2) Z$ .
These symmetries are also reflected in the natural solutions for Eqs.~\eqref{two_stuart_landau_equation_1} and~\eqref{two_stuart_landau_equation_2}: 
The trivial off-state $Z_1 = Z_2 = 0$, the in-phase (IP) oscillatory solution $Z_1 (t) = Z_2 (t) = r \exp(i \hat{\omega} t + \theta)$ with some constant $r$ and $\hat{\omega}$, and arbitrary $\theta$, as well as the anti-phase (AP) solution $Z_1 (t) = - Z_2 (t) = r \exp(i \hat{\omega} t + \theta)$. 

The model presented in Eq.~\eqref{two_stuart_landau_equation_1} and~\eqref{two_stuart_landau_equation_2} can also be seen as the simplest case of a mean-field or globally coupled network of Stuart-Landau oscillators. This system has been studied extensively with respect to chaos in the large oscillator limit \cite{HAK92, NAK93, NAK94a, KU15}. There, cluster states with differing amplitudes have also been reported for mean-field coupled networks. Following Ref.~\cite{HAK92}, these states can be described with equations similar to Eq.~\eqref{two_stuart_landau_equation_1} and~\eqref{two_stuart_landau_equation_2}, but different coupling weights given by the asymmetric sizes of clusters, also see Appendix~D for details. Our focus here is on the symmetric case and in contrast to the existing literature, we find and analytically describe stable symmetry-broken amplitude- and phase-locked solutions. The works of Aronson \textit{et~al.}\cite{ARO90} deserve special mention, as they study the bifurcation structure of two coupled Hopf-normal forms in great detail. However, due to their choice of parameters (coupling shear to frequency), they claim that 'symmetric solutions are the only stable phase-locked solutions' for a symmetric setup. While we agree with their general analysis, we can clearly refute this particular claim for the cases shown in the paper here.

Fig. \ref{2d_plot_multi} shows the results of numerical simulation for varying $\lambda$ and the coupling phase $\phi$ for increasing values of the shear parameter from a)~$\mathbf{Im}(\gamma) = 0.0$ to d)~$\mathbf{Im}(\gamma) = 0.5$). Here, we prepare the system close to one of the three solutions mentioned above, let the system evolve and record the eventually reached stable state. Thus we numerically  obtain an approximate map of stability for these solutions. We also tried random initial conditions but could not find any additional stable states. As can be seen from the colored regions in Fig.~\ref{2d_plot_multi}~a), the different solutions have different regions of stability. Stable in-phase solutions (light peach) are centered around $\phi \approx 0$, while anti-phase solutions (bright blue) appear for $\phi \approx \pi$. There is a substantial overlap between in-phase and anti-phase solutions, indicated by the purple region of multistability in Fig.~\ref{2d_plot_multi}~a). The off-state (white) is only stable for sufficiently small $\lambda$. These types of solutions are very typical of coupled oscillatory systems, and are especially apparent in coupled lasers \cite{YAN04c, CLE14, BOE16, ROE16}, the beginnings of which go back to the early days of laser theory \cite{SPE72}.

The lower half of Fig. \ref{2d_plot_multi} shows the corresponding bifurcation lines of which the relevant parts are reproduced in the top-panel. Bifurcations can be derived analytically with standard approaches for all important stability boundaries, of which the details can be found in the Appendix. The main bifurcations are the Andronov-Hopf bifurcations creating stable in-phase (H$_{IP}$, black) and anti-phase (H$_{AP}$, grey) limit cycles: 
\begin{align}        
\lambda_{H(IP)} &= 0, \label{HIP}\\
\lambda_{H(AP)} &= 2 \kappa \cos \phi. \label{HAP}
\end{align}
For $\mathbf{Re}(\gamma) < 0$, the case shown in Fig.~\ref{2d_plot_multi}, these are supercritical. Two Hopf-Hopf-points (black circles in Fig.~\ref{2d_plot_multi}) situated at $\phi = \pi/2, 3 \pi/2$ give birth to two additional secondary Andronov-Hopf or Torus-Bifurcations occurring for the in-phase (T$_{IP}$, dark blue) and anti-phase (T$_{AP}$, orange) limit cycle, respectively: 
\begin{align}        
\lambda_{T(IP)} &=  - 2 \kappa \cos \phi  \label{TIP}\\
\lambda_{T(AP)} &= 4 \kappa \cos\phi. \label{TAP}
\end{align}
Additionally, a pitchfork of limit cycles also limits the region of stability (P$_{IP}$, violet, and P$_{AP}$, brown):
\begin{align}        
\lambda_{P(IP)} &=-\frac{\kappa  \mathbf{Re}(\gamma)}{\mathbf{Im}(\gamma) \sin \phi +  \mathbf{Re}(\gamma)\cos \phi}  \label{PIP}\\
\lambda_{P(AP)} &=\frac{ \kappa \mathbf{Re}(\gamma)}{\mathbf{Im}(\gamma) \sin \phi  + \mathbf{Re}(\gamma) \cos \phi} + 2 \kappa \cos \phi. \label{PAP}
\end{align}
The pitchfork bifurcation of the in-phase solution (violet line in Fig.~\ref{2d_plot_multi}) has also been identified with the Benjamin-Feir instability in the case of large networks of mean-field coupled Stuart-Landau oscillators (cf. Ref.~\cite{HAK92}, and Appendix Sec.~D)).

The symmetric picture of Fig.~\ref{2d_plot_multi}~a) gets distorted once we set $\textbf{Im}(\gamma) \not= 0$. This is the more general case describing a 'soft' spring, i.e. a spring that changes its frequency depending on the amplitude of oscillations. The actual sign of $\textbf{Im}(\gamma)$ does not influence our results except for a change of sign in $\phi$. 
The results for non-zero $\textbf{Im}(\gamma) \not= 0$ can be seen in Fig.~\ref{2d_plot_multi}~b)-d): The Andronov-Hopf-bifurcations (H$_{IP}$ and H$_{AP}$) and secondary Andronov-Hopf (Torus) bifurcations (T$_{IP}$ and T$_{AP}$) have not changed, as expected from the analytic formulas of Eqs.~\eqref{HIP}-\eqref{TAP}. However, the pitchfork bifurcations of limit cycles (P$_{IP}$ and P$_{AP}$) have shifted and distorted significantly (brown and purple lines in the lower panels of Fig.~\ref{2d_plot_multi}). The bifurcations are no longer mirrored along $\phi = \pi$. At some value of $\textbf{Im}(\gamma)$ the Torus (T) and pitchfork (P) bifurcation cross each other (compare Fig.~\ref{2d_plot_multi}~b) and~c)) and then some regions no longer show any stable solutions previously discussed (grey and dark green areas). Most of these regions contain higher-order limit cycles and quasi-periodic behavior, which can be in-phase, anti-phase and even phase-unbounded. The regions marked in green contain a new solution, which we will refer to as 'symmetry-broken amplitude- and phase-locking' (SBL).


\section{Symmetry-broken amplitude- and phase-locking}

\subsection{Description}

\begin{figure*}
        \includegraphics[width=1.00\linewidth]{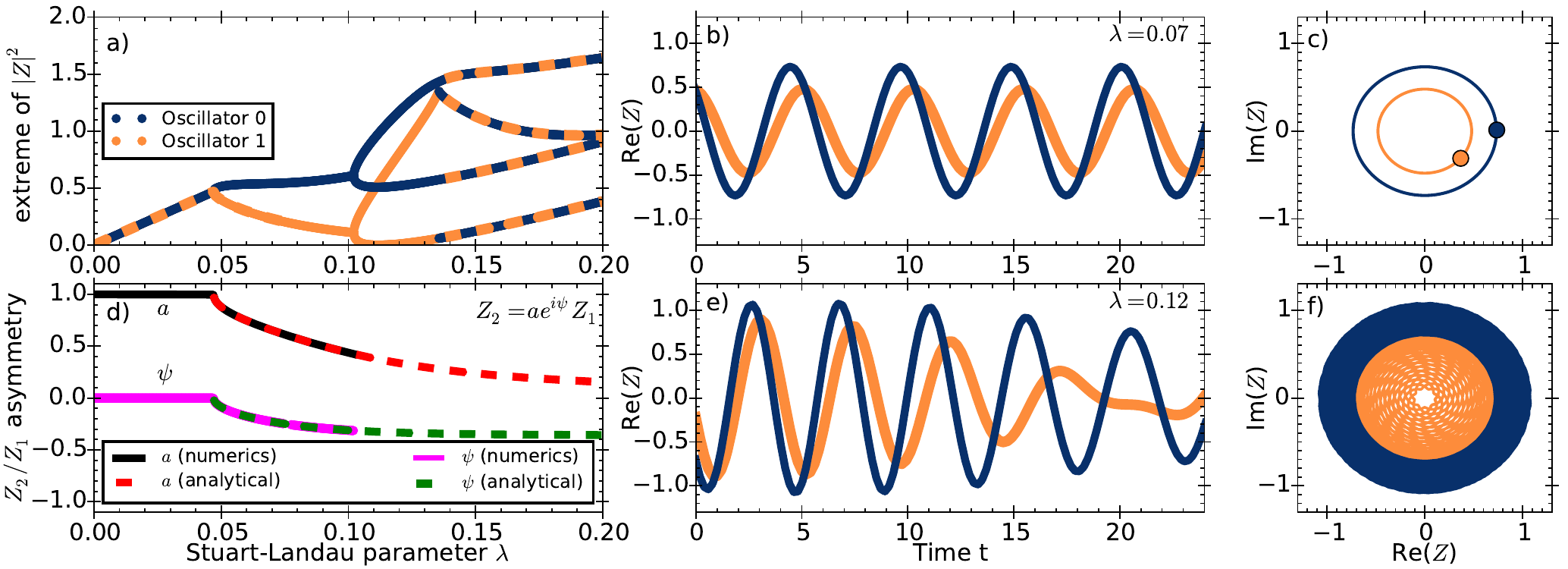}
        \caption{a) Numerical linescan showing the extrema of two coupled Stuart-Landau-Oscillators (blue and yellow). Where blue and yellow diverge, oscillators are symmetry-broken and possess different extrema. d) Numerically evaluated amplitude ratio $a$ (black) and phase lag parameter $\psi$ (purple) as defined by $Z_2 = a \exp(\ii \psi) Z_1$. Analytically derived predictions for $a$ and $\psi$ from Eq.~\eqref{Supp_Eq_7} and~\eqref{Supp_Eq_8} (Appendix) are shown in red and green. Timeseries showing the real part of $Z_1$  and $Z_2$ for Eqs.~\eqref{two_stuart_landau_equation_1} and~\eqref{two_stuart_landau_equation_2} after a transient time of 10000 time units are shown in b) for $\lambda = 0.07$ and e) for $\lambda = 0.12$. Panel b) is an example for symmetry-broken amplitude- and phase-locking. Panels c) and f) show phase-space portraits of b) and e), respectively. Parameters: $\mathbf{Re}(\gamma) = -0.1$, $\mathbf{Im}(\gamma) = 0.5$, $\omega = 1$, $\kappa = 0.1$, $\phi = 0.2 \pi $, $\lambda = 0.12$}
        \label{timeseries_and_linescan}
\end{figure*}

The regions marked in green in Fig.~\ref{2d_plot_multi}~c)-d) contain an additional type of stable synchronization, which we had not anticipated to find. These symmetry-broken amplitude- and phase-locking states are characterized by harmonic, regular oscillations with the same frequency for both oscillators, similar to the in-phase and anti-phase solutions, but the oscillators differ in amplitude and phase. Hence, in the phase-space of $\mathbf{Re}(Z)$ and $\mathbf{Im}(Z)$ these solutions appear as two discrete circular limit cycles for $Z_1$ and $Z_2$ as shown in Fig.~\ref{timeseries_and_linescan}~c). In a co-moving frame these states are fixed points, which distinguishes them from the symmetry-broken intensity oscillations found for a small network of lasers in Ref.~\cite{ROE16}. While a lot has been done on synchronization of Stuart-Landau  oscillators \cite{PIK01}, we could not find any references to these 'symmetry-broken amplitude- and phase-locking' states in the literature, except for Aronson \textit{et al.}, who found them to be always unstable \cite{ARO90}. And even there, the authors did not calculate the solutions in detail, nor did they report any global properties of these states, as their focus was on non-identical oscillators. Indeed, as seen by the numerical results in Fig.~\ref{2d_plot_multi}, these states only become visible for sufficiently high $\textbf{Im}(\gamma)$, which may explain their absence from the literature. They do however exist for most of the parameter plain of $\lambda > 0$, as we will show. They are locally stable where a supercritical pitchfork bifurcation bounds the in-phase and anti-phase regions of stability. 

The symmetry-broken amplitude- and phase-coupling states present an interesting case of how symmetry-breaking bifurcations can be used to generate bistable switches. Due to the symmetries of Eq.~\eqref{two_stuart_landau_equation_1} and~\eqref{two_stuart_landau_equation_2}, it is impossible to predict which oscillator will have the larger amplitude. While the $\mathbb{Z}_2 $-exchange symmetry is broken within a single example of these states, the symmetrized version must always exist according to the \textit{Equivariant Branching Lemma} \cite{CRA91, HOY06}. This creates a natural multistability between these two solutions, as they must share the same regions of stability to fulfill the symmetries of the underlying system. From the perspective of one of the oscillators, this means that two stable harmonic limit cycles coexist simultaneously. When used as switches, one usually prefers 'nice' and 'well-behaved' solutions. The symmetry-broken amplitude- and phase-locking solutions preserve the $S^1$-symmetry of the system, i.e. they have no amplitude oscillations, and therefore represent a perfect example for use in applications.



Figure~\ref{timeseries_and_linescan}~a) shows the extrema of $|Z|^2$ versus $\lambda$ for $\phi = 0.2 \pi$. This is a representative linescan for the creation and destabilization of symmetry-broken amplitude- and phase-locking states. First, the usual onset of harmonic oscillations at $\lambda = 0$ can be seen by the existence of a single maximum in $|Z|^2$ for $\lambda > 0$. For this the value of coupling phase $\phi$, only the in-phase synchronized solution becomes stable and $|Z|^2$ is identical for both oscillators. However at $\lambda \gtrsim 0.05$, the maxima of both oscillators begin to differ, which indicates the symmetry-breaking in both the amplitude and the phase, without any asymmetry in the oscillator parameters or coupling terms. The underlying bifurcation is a pitchfork of limit cycles, which breaks the underlying exchange symmetry. The symmetry-breaking nature of pitchfork bifurcations is well-established \cite{HEI10}, but noticeably, here the harmonic oscillations are nevertheless preserved. This also distinguishes these states from the known phenomenon of oscillation death \cite{KOS13, ZAK13}, which stabilizes asymmetric fixed points and employs a symmetry-breaking coupling. A representative timeseries slice (after a transient time of 10000 time units) for $\Re (Z)$ is shown in Fig.~\ref{timeseries_and_linescan}~b) and a phase space representation of $Z$ in Fig.~\ref{timeseries_and_linescan}~c).
For $\lambda \gtrsim 0.1$, we find an additional bifurcation, which we so far cannot describe analytically. However, numerics strongly imply that it is a secondary Andronov-Hopf, leading to amplitude oscillations on top of the symmetry-broken amplitude- and phase-locking state. A sample time series (after a transient time of 10000 time units) is shown in Fig.~\ref{timeseries_and_linescan}~e). After the secondary Andronov-Hopf, oscillations are no longer harmonic and thus the phase space projection in Fig.~\ref{timeseries_and_linescan}~f) is filling out an area. Resymmetrization of both oscillators happens, in this example but not always, after an additional period-doubling for $\lambda \gtrsim 0.13$. We have not investigated this bifurcation in greater detail, but it seems to be a symmetry-breaking period-halfing bifurcation in the reverse direction.

%
%
%


\subsection{Analytical description for the general case}

\begin{figure*}
        \includegraphics[width=1.00\linewidth]{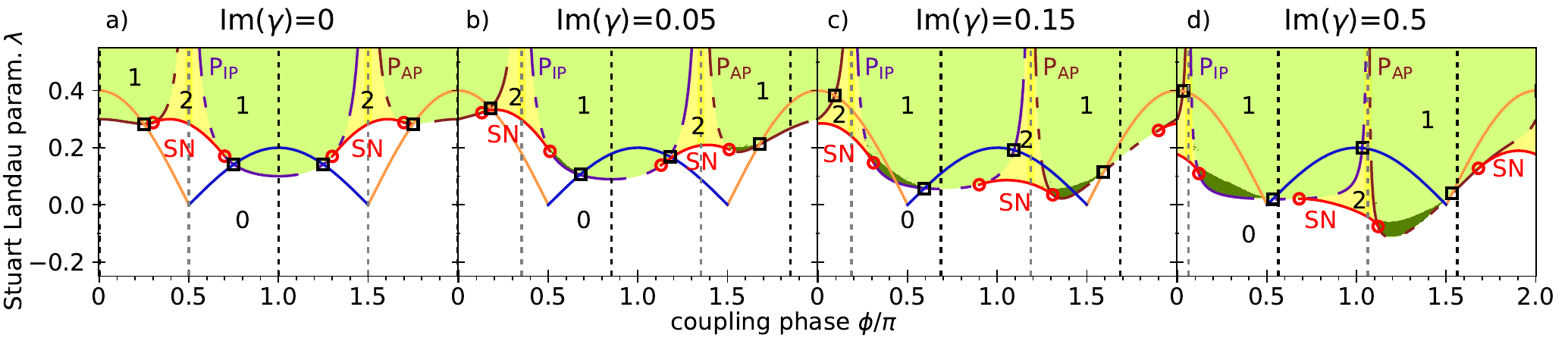}
        \caption{Analytically derived existence of the symmetry-broken amplitude- and phase-locking (SBL) states for different coupling phases $\phi$ and Stuart-Landau parameter $\lambda$ shown for increasing $\mathbf{Im} \gamma$ in a)-d). The number indicates the number of different branches of symmetry-broken amplitude- and phase-locking states (excluding the exchange symmetry). Pitchfork bifurcation lines are given according to Eq.~\eqref{PIP} and~\eqref{PAP} (dashed lines), torus bifurcations lines according to Eq.~\eqref{TIP} and~\eqref{TAP} (solid blue and orange lines), and saddle-node lines as given by Eq.~\eqref{SN} (solid red lines). Saddle-node-pitchfork points are indicated by red circles, pitchfork-torus points by black squares. Vertical black dotted lines show where symmetry-broken amplitude- and phase-locking solutions with $\psi = 0$ and $\psi = \pi$ exist, grey dotted lines show $\psi = \pi/2$ and $\psi = 3 \pi/2$. Dark green areas show the numerically found regions of stable amplitude- and phase-locking, light green and yellow the existence. Parameters: $\mathbf{Re}(\gamma) = -0.1$, $\mathbf{Im}(\gamma)$ as shown atop, $\omega = 1$, $\kappa = 0.1$}
        \label{asymmmetric_states_fig}
\end{figure*}

After the numerical indications for the existence of the symmetry-broken amplitude- and phase-locking states, we will now show how to derive an analytical formula to describe these states. Let $s \in \mathbb{C}$  be the complex factor describing the relationship of $Z_1$ and $Z_2$ at all times:
\begin{align}        
	Z_2 &= s Z_1 =: s Z,
	\label{s_ansatz}
\end{align}
which we can insert into Eqs.~\eqref{two_stuart_landau_equation_1} and~\eqref{two_stuart_landau_equation_2}. 
We use the transformations $\Lambda = \lambda - \textbf{Re}(\sigma)$, $\tilde{\omega} = \omega - \textbf{Im}(\sigma)$ and the complex coupling strength $\sigma = \kappa \exp \ii \phi$. We are specifically looking for states with harmonic oscillations and can therefore set $\dot{Z} = \ii \tilde{\Omega} Z$, with some unknown $\tilde{\Omega}$. After dividing by $Z$ (assuming $Z \not = 0$) we obtain
\begin{align}
0 &= \left(\Lambda+\ii \Omega + \gamma\,|Z|^{2}\right) + \sigma s \label{sl_derivation_ref1}\\
0  &= \left(\Lambda+\ii \Omega + \gamma\,|s|^{2} |Z|^{2}\right) s  +\sigma \label{sl_derivation_ref2},
\end{align}
with $\Omega = \tilde{\omega} - \tilde{\Omega}$. 
The two complex Eqs.~\eqref{sl_derivation_ref1} and~\eqref{sl_derivation_ref2} contain the unknown quantities $|Z |^2$, $\Omega$ and $s = \mathbf{Re} (s) + \ii \mathbf{Im} (s)$. They can be used to obtain expressions for these quantities as a function of the other parameters ($\lambda$, $\phi$, $\gamma$). Using polar description:
\begin{align}
s &= a \exp (\ii \psi),
\end{align}
 one can obtain a short equation for the relationship of $a$ and $\psi$:
\begin{align}        
	a &= \sqrt{\frac{\mathbf{Im} (\gamma)  \cos \left(\phi - \psi \right) - \mathbf{Re} (\gamma) \sin \left(\phi - \psi \right)}{\mathbf{Im} (\gamma)  \cos \left(\phi + \psi \right) - \mathbf{Re} (\gamma) \sin \left(\phi + \psi \right)}}.
	\label{a_of_psi_equation}
\end{align}
This extends the work of Aronson \text{et al.}\cite{ARO90}, who described $\psi = 0$ for $\mathbf{Im} (\gamma) = 0$. However, one arrives at transcendental equations if one tries to solve Eqs.~\eqref{sl_derivation_ref1} and~\eqref{sl_derivation_ref2} completely for $a$ and $\psi$ using polar coordinates. We therefore used $s = \mathbf{Re}(s) + \mathbf{Im}(s)$ and split each equation into real and imaginary part. The details of the derivation and full solutions are shown in Appendix B. Because they are the solutions to a quartic equation, the expressions become somewhat lengthy, see Appendix B Eq.~\eqref{Supp_Eq_7} and~\eqref{Supp_Eq_8}. Additionally, one must be careful to exclude spurious solutions, which can be introduced. Nevertheless, we did obtain complete analytical formulas, that do not rely on any implicit solution methods. With these solutions, we can show that every phase-lag $\psi$ can be reached with a broken symmetry even the 'symmetric phases' of $\psi = 0, \pi$, i.e identical phases but different amplitudes. However, we never find a state with $a = 1$ and $\psi \not = 0, \pi$, i.e. identical amplitudes but an asymmetric phase-relationship. Fig.~\ref{timeseries_and_linescan}~d) compares the numerically obtained phase and amplitude relationship $Z_2 = a \exp (\ii \psi) Z_1$ (blue and pink), with the analytic expressions for $a$ (pink) and $\psi$ (green). As can be seen in Fig.~\ref{timeseries_and_linescan}~d), the agreement is perfect, as no approximations were made in the derivation of the analytical formulas. The analytical forms can now be used to derive additional bifurcation conditions for the symmetry-broken amplitude- and phase-locking states. 

The bounding bifurcations and regions of existence are shown in Fig.~\ref{asymmmetric_states_fig} a)-d) for the same values of $\mathbf{Im} (\gamma)$ as in Fig.~\ref{2d_plot_multi} a)-d). We have used the analytically derived bifurcations to calculate the regions of existence for symmetry-broken amplitude- and phase-locking states. The number of qualitatively different co-existing branches is indicated with the color scheme and numbers, where green denotes the existence of one branch of symmetry-broken amplitude- and phase-locking states, and yellow the existence of two such branches. We have not counted the symmetrized versions as independent solutions. As can be seen, for large enough $\lambda$ such states always exist.

The pitchfork of limit cycles described in Sec.~\ref{sec_bif} and shown in Eq.~\eqref{PIP} and~\eqref{PAP}, which bounds the stability region of the in-phase and anti-phase orbits and was already identified in Ref.~\cite{HAK92} as the Benjamin-Feir-instability, at the same time generates the symmetry-broken amplitude- and phase-locking states. It is shown as P$_{IP}$ and P$_{AP}$ in Fig.~\ref{2d_plot_multi} and Fig.~\ref{asymmmetric_states_fig}. In addition to the pitchfork of limit cycles, symmetry-broken amplitude- and phase-locking can also arise in a saddle-node bifurcation described by:
\begin{align}
\lambda_{SN} =& \frac{\sqrt{8 \kappa^2 (c \cos \phi - \sin \phi)(c^2 \sin \phi + 2 c \cos \phi - \sin \phi)}}{c^2 + 1} |\kappa \sin \phi| \nonumber \\
&- \kappa \cos \phi, \label{SN}
\end{align}
with $c = \mathbf{Im} (\gamma)/\mathbf{Re} (\gamma)$. This saddle node bifurcation gives rise to two branches of symmetry-broken amplitude- and phase-locking states.  One of these branches connects to the pitchfork of limit-cycles, which is subcritical for all values of $\phi$ for which the saddle-node exists and connected via the unstable branch of the saddle-node bifurcation. Consequentially, where the saddle-node bifurcation collides with the pitchfork of limit cycles, the latter turns from supercritical to subcritical. This is indicated by the red circles in Fig.~\ref{asymmmetric_states_fig}. This saddle-node bifurcation arises as a necessary condition while calculation $a$ and $\psi$, of which the details are shown in Appendix B. To our knowledge, its existence was not reported before and completes the boundaries of the region where symmetry-broken amplitude- and phase-locking states exist.

Knowing all the boundaries, it is therefore possible to describe the local stability of the symmetry-broken amplitude- and phase-locking states. Where the pitchfork is supercritical, the stability of in-phase or anti-phase solutions is transferred to the newly created symmetry-broken amplitude- and phase-locking state. Thus, these states are born stable along the parts of the pitchfork lines (P$_{IP}$ and P$_{AP}$) that lie between the saddle-node-pitchfork-point (red circles in Fig.~\ref{asymmmetric_states_fig}) and the pitchfork-torus-point (black squares in Fig.~\ref{asymmmetric_states_fig}), i.e. where the pitchfork is supercritical and bounds the regions of stability of the in-phase and anti-phase solutions as shown in Fig.~\ref{2d_plot_multi}. Especially of note is that initially, these states appear on both sides of the pitchfork, but for increasing $\mathbf{Im} (\gamma)$ the situation becomes asymmetric as the saddle-node-pitchfork-point and the pitchfork-torus-point cross each other (compare circles and squares in Fig.~\ref{asymmmetric_states_fig} a) and b), first the red circles are surrounded by black squares, then it becomes an alternating pattern).

\subsection{Symmetry-broken amplitude- and phase-locking for the special case of phase-lag $\psi = 0,\pi$ and $\psi = \pi/2,\ 3\pi/2$}
\label{sec:special_cases}

While generically the symmetry-broken amplitude- and phase-locking combines both differences in the amplitude as well as in the phase, we can also find parameters where only the amplitude differs, and the phase is trivially fixed, i.e $Z_2 = a \exp (\ii \psi) Z_1$ with either $\psi = 0$ or $\psi = \pi$.
The full calculations can be found in Appendix~C.
The condition for this special case is given by
\begin{equation}
\frac{\mathbf{Im} (\gamma)}{\mathbf{Re} (\gamma)}= \frac{\mathbf{Im} (\sigma)}{\mathbf{Re} (\sigma)},
\label{special_case_cond_1}
\end{equation}
where $\sigma = k \exp (i\phi)$ was used. This condition corresponds to the black vertical dotted lines in Fig.~\ref{asymmmetric_states_fig}, and they appear at different coupling phases depending on the shear parameter $\mathbf{Im}( \gamma)$. Nevertheless, they always appear in the center of the regions bounded by the in-phase pitchfork (PIP) and anti-phase pitchfork (PAP) bifurcation lines. 
Along these vertical lines we can calculate the amplitude relation $a$ explicitly, depending on the bifurcation parameter $\lambda$.
Bifurcating from the in-phase solution, we find the two branches
\begin{equation}
a_{\pm}(\lambda)=\frac{\lambda-\mathbf{Re} (\sigma) \pm \sqrt{-3 \mathbf{Re} (\sigma)^2 - 
  2 \mathbf{Re} (\sigma) \lambda+ \lambda^2}}{2 \mathbf{Re} (\sigma)}
  \label{special_case_branches_1}
\end{equation}
Note, that $a_+ = 1/a_-$. On one of these branches, $a$ approaches zero asymptotically, while the other branch grows linearly with $\lambda$. 
The branches bifurcating from the anti-phase solution have almost the same form:
\begin{equation}
a_{\pm}(\lambda)=\frac{\mathbf{Re} (\sigma) - \lambda \pm \sqrt{-3 \mathbf{Re} (\sigma)^2 - 
  2 \mathbf{Re} (\sigma) \lambda+ \lambda^2}}{2 \mathbf{Re} (\sigma)}
   \label{special_case_branches_2} 
\end{equation}
The second case $\psi = \pi/2,\ 3\pi/2$ can be treated analogously. Here we find
the condition
\begin{equation}
\frac{\mathbf{Im} (\gamma)}{\mathbf{Re} (\gamma)}= -\frac{\mathbf{Re} (\sigma)}{\mathbf{Im} (\sigma)}.
\label{special_case_cond_2}
\end{equation}
Hence a phase shift of $\psi = \pi/2,\ 3\pi/2$ occurs exactly for coupling phases $\phi$ where there is no pitchfork bifurcation and the symmetry-broken amplitude-and phase-locked states only emerge from saddle-node bifurcations, see the grey dotted vertical lines in Fig.~\ref{asymmmetric_states_fig}.
The full equations $a(\lambda)$ for this case can be found in Appendix~C.

The conditions shown in Eq.~\eqref{special_case_cond_1} and~\eqref{special_case_cond_2} describe the only coupling-phases $\phi$ for which $\psi$ does not change while increasing $\lambda$, only the amplitude ratio $a$ changes. Therefore, the four vertical dotted lines in Fig.~\ref{asymmmetric_states_fig} form a sort of 'organizing skeleton' at which the symmetry-broken amplitude- and phase-locking states are pinned.

%
%
%


\section{Discussion}

The symmetry-broken amplitude- and phase-locking presented in this paper may be related to what people generally refer to as 'phase-lag synchronization'. But while 'phase-lag synchronization' usually refers to synchronization of noisy or chaotic systems, the states found here are absolutely regular and thus of a somewhat different nature. Similarly, these symmetry-broken amplitude- and phase-locking states may be seen as a peculiar case of a splay-state. However, in splay-states one usually finds a symmetric distribution of phase-lags between clusters, which is not the case here. 

There is also the argument to be made that these states are connected to cluster-solutions found in large mean-field coupled oscillator networks. The main bifurcations of this system have already been known and can also be derived from more general $N$-oscillator networks (see the Appendix D)). Given the important position of the Stuart-Landau model in the field of nonlinear dynamics, this system has naturally been studied thoroughly in the past. When comparing with the results of Ref.~\cite{KU15} a similarity between cluster states and symmetry-broken amplitude- and phase-locking states is apparent. However, we still believe the connection between cluster states and the symmetry-broken amplitude and phase-locking states to be more complex than that. While both cluster states and the states presented in this paper are born in symmetry-breaking bifurcations, their stability differs in a crucial way: In the case of large $N$, the cluster states as described in the literature appear to be dominated by the internal stability of clusters of different sizes. Once the sizes of such clusters are known, the dynamics can be calculated using a modified version of Eq.~\eqref{two_stuart_landau_equation_1} and~\eqref{two_stuart_landau_equation_2}, where one adds a weight in the coupling term. However, these weighted coupled Stuart-Landau equations will no longer possess the $\mathbb{Z}_2$-symmetry and hence differing amplitudes are to be expected for the clusters. Until investigated more clearly, it is not clear whether a solution with 2 clusters of the same size but different amplitudes is stable in a system with large $N$ and no detailed description of solutions breaking only the $\mathbb{Z}_{2}$-symmetry is given in the literature for a case of two clusters.
In our case here, with only two oscillators, each of them exhibiting a different phase, applying the term 'cluster' seems unsuited. From a applications perspective systems with many oscillators forming clusters both exhibit unwanted complexity and unneeded degeneracy, with solutions with amplitude-oscillations potentially coexisting with fixed-amplitude states (cf. Fig.~4. in Ref.~\cite{KU15}).

Schmidt and Krischer \cite{SCH14n} investigated a network of Stuart-Landau oscillators with a non-linear coupling. The system contains an additional conservation law, allowing for analytical solutions despite the more complex equations of motion. They report the existence and stability regions of "amplitude clusters" and "modulated amplitude clusters", which are likely analog to the states reported in this work. This possibly suggests, that the symmetry-broken amplitude- and phase-locking states are not tied to a specific form of coupling.

As discussed in Sec.~\ref{intro}, the applicability of our results to a wide range of systems is possible, due to the general nature of the Stuart-Landau equation. Arguably the most important example are laser systems, for which the corresponding Lang-Kobayashi equations (excluding the delay term) are very similar to a Stuart-Landau system as shown in  Eq.~\eqref{two_stuart_landau_equation_1}, and given by:
\begin{align}
    \frac{dE}{dt}=&(1+i\alpha) E N \label{E-equation}\\
    \frac{dN}{dt}=&\frac{1}{T}(p-N-(1+2N)|E|^2) , \label{N-equation}
\end{align}
where $E$ is the complex electric field variable, $N$ the normalized excess carrier density, $T$ the electron-photon lifetime ratio, $\alpha$ is the amplitude-phase-coupling and $p$ the pump current. Indeed, for two instantaneously coupled lasers, a similar state was already mentioned in Ref.~\cite{YAN04c}, albeit it was never found to be stable. For the case with delayed coupling Ref.~\cite{JAV03} presents a similar solution. Recently, the authors of Ref.~\cite{CLE14} reported stable solutions analogous to the symmetry-broken amplitude- and phase-locking states found in this paper and called them 'single-color symmetry-broken states'. These were found using numerical simulations and path continuation methods for $\tau =0$, which was extended to $\tau \not = 0 $  in Ref.~\cite{SEI17}. In fact, the amplitude-modulated version of the symmetry-broken amplitude- and phase-locking states shown in Fig.~\ref{timeseries_and_linescan}~e) is possibly connected to the 'two-color symmetry-broken states' of Ref.~\cite{CLE14}. However, in comparison to the Stuart-Landau case presented here, no analytical description for the laser equations seems to be known, as only implicit transcendental equations are derived. Additionally, the global picture of bifurcations already seems to differ in some fundamental aspects, as there are only miniscule bistable regions between in-phase and anti-phase solutions at small coupling strengths, while for large coupling strength stable in-phase and anti-phase solutions are bounded by two tightly separated torus-bifurcations. Another discussion of symmetry-broken laser states can be found in Ref.~\cite{KOM17}. As laser dynamics is a vast topic with large diversity of effects \cite{LIN15a, ROE16}, the interplay between symmetry-broken amplitude- and phase-locking states and more complex dynamics can be explored \cite{SIV17}.

In our parameter study, we found the imaginary part of the nonlinearity parameter $\gamma$ to be important. While the sign does not change our results, except for a mirroring along $\phi = \pi$, the size of $\mathbf{Im}(\gamma)$ determines the extend of the region of stable symmetry-broken amplitude- and phase-locking. Additionally, as the system can be rescaled with respect to $\mathbf{Re}(\gamma)$, it is best to consider the fraction $\mathbf{Im}(\gamma)/\mathbf{Re}(\gamma)$. This corresponds to the well known amplitude-phase-coupling factor $\alpha$ in the laser system seen in Eq.~\eqref{E-equation}. The parameters used in this work would translate to $\alpha = 5$ for $\mathbf{Im}(\gamma) = 0.5$. Similarly, $\lambda$ can be related to the driving pump current $p$ of a laser system. We therefore expect symmetry-broken amplitude- and phase-locking to be more prevalent in high-$\alpha$ lasers sufficiently above threshold.

In coupled nonlinear tunnel diodes, the authors of Ref.~\cite{HEI10} describe 'skew orbits' and symmetry-broken stable states. While the used system there has no $S^1$-symmetry, the results are still underlining how a pitchfork bifurcation can result in symmetry-broken solutions.

One important result of our work is the existence of bistable and tristable regions for coupled symmetric nonlinear oscillators. These can be used as switches, and in fact this has been demonstrated numerically for the laser system in Ref.~\cite{CLE14}. We expect the same methods to be applicable to all nonlinear oscillators coupled in a scheme similar to the one used here.

Lastly, as we can show the existence of symmetry-broken amplitude- and phase-locking states for any coupling-phase $\phi$ given a sufficiently large $\lambda$ (cf. Fig.~\ref{asymmmetric_states_fig}), this opens the possibility to stabilize these states using feedback methods.

\section{Conclusion}

To conclude, we found symmetry-broken amplitude- and phase-locking in a system of two coupled identical Stuart-Landau oscillators. These states are characterized by harmonic oscillations of both oscillators with different amplitudes and phases. They break the underlying exchange symmetry of the system, without destroying the harmonic oscillations. We calculated the bifurcations responsible for their creation and can analytically derive expressions for the phase-lag $\psi$ and amplitude-relationship $a$ of both oscillators. Additionally, we presented a parameter study, showing the increased stability of these solutions with increasing shear $\mathbf{Im}(\gamma)$. There is a strong connection to previously found results in laser systems and we expect these results to be applicable to a wide range of coupled nonlinear oscillators. The nature of these states makes them perfect as switches, owing to the natural multistability.

\section{Acknowledgements}

This work was supported by the Deutsche Forschungsgemeinschaft (DFG) within the framework of the SFB910.  We thank J.~Sawicki, B.~Fiedler, A.~Zakharova, E.~Sch\"oll, A.~L\'opez Nieto, B.~Lingnau, and L.~Jaurigue for their support, fruitful discussions and helpful remarks.

\section*{Appendix}

Here we present the detailed derivations of the analytical results of the main paper. These technical details will be most useful to readers trying to transfer our results to other systems. The Appendix first derives the standard bifurcations (Eq.~\eqref{HIP} to~\eqref{PAP}) in App.~A. We then derive the analytical form of the symmetry-broken amplitude- and phase-coupling states in App.~B, with the special cases of $\psi$ covered in App.~C. Lastly, we compare and discuss these results with previous publications on globally coupled Stuart-Landau networks in App.~D.

\subsection{Bifurcations}

This section describes how to derive the bifurcation lines shown in Fig.~\ref{2d_plot_multi}. For the system as described in Eqs.~\eqref{two_stuart_landau_equation_1}-\eqref{two_stuart_landau_equation_2} an Andronov-Hopf bifurcation occurs at $\lambda=0$, as it is the normal form of such a bifurcation. There, the oscillators are in-phase ($Z_1 = Z_2$). The phase velocity of this orbit is given by 
\begin{equation}
\Omega = \omega + \mathbf{Im}(\gamma) \,r^{2} = \omega - \lambda \mathbf{Im}(\gamma)/\mathbf{Re}(\gamma),
\end{equation}
where the radius $r$ grows with 
\begin{equation}
r= \sqrt{- \lambda/\mathbf{Re}(\gamma)}.
\end{equation}

By moving to a co-rotating frame $\hat{Z} = \exp(-\ii \Omega t) Z$ we can eliminate the rotation from the system. All oscillations with fixed amplitude and frequency $\Omega$ become fixed points in the new variable $\hat{Z}$ (dropping the hat from now on). In this new system the stability of fixed points can be evaluated by linear stability analysis. The stability of a fixed point ($Z_1 = x_1 + \ii y_1, Z_2 = x_2 + \ii y_2$) is calculated from the Jacobian of the linearized system given by
\begin{align}
\begin{pmatrix}
-C + J(Z_1) & C\\
C & -C + J(Z_2)\\
\end{pmatrix}
\end{align}
with $C$ given by
\begin{align}
\begin{pmatrix}
\mathbf{Re}(\sigma)	&-\mathbf{Im}(\sigma)\\
\mathbf{Im}(\sigma)	&\mathbf{Re}(\sigma)
\end{pmatrix}.
\end{align}
Here, $\sigma$ is the complex coupling strength $\sigma = \kappa \exp(\ii \phi)$. $J(Z_{n})$ is defined as 
\begin{widetext}
\begin{align}
\begin{pmatrix}
2 \mathbf{Im}(\gamma) x_{n} y_{n}+\mathbf{Re}(\gamma) \left(3 x_{n}^2+y_{n}^2\right)+\lambda &	2 \mathbf{Re}(\gamma) x_{n} y_{n}-\mathbf{Im}(\gamma) \left(x_{n}^2+3 y_{n}^2\right)+\Omega-1	\\
2 \mathbf{Re}(\gamma) x_{n} y_{n}+\mathbf{Im}(\gamma) \left(3 x_{n}^2+y_{n}^2\right)-\Omega+1 & 2\mathbf{Im}(\gamma) x_n y_n+\mathbf{Re}(\gamma) \left(x_{n}^2+3 y_{n}^2\right)+\lambda 
\end{pmatrix}
\end{align}
\end{widetext}

Inserting the values for the in-phase solution $Z_1 = Z_2$,  we can calculate the characteristic equation:
\begin{align}
0 =\eta (\eta + 2 \lambda)\Big( & 4 \mathbf{Im}(\sigma)^2  +  4 \frac{\mathbf{Im}(\sigma)}{ \mathbf{Re}(\gamma)} \mathbf{Im}(\sigma) \nonumber \\   
   & + (2 \mathbf{Re}(\sigma) + 2 \eta) (2 \mathbf{Re}(\sigma) +  \eta + 3 \lambda)\Big),
\label{char_polyn}
\end{align}
where $\eta$ is the eigenvalue we are interested in.

Eq.~\eqref{char_polyn} always has the solution $\eta = 0$, corresponding to the trivial Floquet exponent. The next factor yields a zero Flocquet exponent for $\lambda = 0$, corresponding to the Andronov-Hopf bifurcation generating the in-phase solution ($H_{IP}$ in Fig.~\ref{2d_plot_multi}). The final factor given by
\begin{align}
4  & \mathbf{Im}(\sigma)^2  + 
 4 \frac{ \mathbf{Im}(\gamma)}{  \mathbf{Re}(\gamma)}  \mathbf{Im}(\sigma) \lambda \nonumber \\ 
& + (2  \mathbf{Re}(\sigma) + 2 \eta) (2  \mathbf{Re}(\sigma) +  \eta + 
 3 \lambda)
 \label{last_factor}
\end{align}
always gives pairs of eigenvalues. Setting $\eta = 0$ yields the condition
 \begin{equation}
\lambda_{P(IP)} =-\frac{\kappa  \mathbf{Re}(\gamma)}{
 \mathbf{Im}(\gamma) \sin \phi +  \mathbf{Re}(\gamma)\cos \phi},   
\end{equation}
which corresponds to the in-phase pitchfork of limit cycles ($P_{IP}$ in Fig.~\ref{2d_plot_multi}). Additionally, we can look for eigenvalues with non-vanishing imaginary part, $ \mathbf{Re}(\eta) = 0$ in Eq.~\eqref{last_factor}. We find
\begin{equation}
\lambda_{T(IP)} =-2  \mathbf{Re}(\sigma) = - 2 \kappa \cos \phi.
\end{equation}
corresponding to a secondary Hopf or Torus bifurcation ($T_{IP}$ in Fig.~\ref{2d_plot_multi}).


Starting from the trivial off-state $Z_1 = Z_2 = 0$, another Hopf-bifurcation occurs at $\lambda_{H(AP)} = 2 \kappa \cos \phi$. There, the oscillators are anti-phase ($Z_1 = -Z_2$). The phase velocity of this orbit is given by 
\begin{equation}
\Omega = \omega +  \mathbf{Im}(\gamma)\, r^{2}-2  \mathbf{Im}(\sigma),
\end{equation}
where the radius $r$ is given by 
\begin{equation}
r= \sqrt{\left(2 \kappa \cos(\phi) - \lambda\right)/ \mathbf{Re}(\gamma)}.
\end{equation}
Again we transform the system into the co-rotating frame. The resulting Jacobian is identical to the in-phase case. However, inserting the anti-phase solution yields a different characteristic polynomial:
%
\begin{align}
0 & =  \eta \Big(4  \mathbf{Im}(\sigma)^2 + 
   \frac{ \mathbf{Im}(\gamma)}{ \mathbf{Re}(\gamma)}  \mathbf{Im}(\sigma) (8  \mathbf{Re}(\sigma) - 4 \lambda)  \nonumber \\
    & + (2  \mathbf{Re}(\sigma) - \eta) 
    (6  \mathbf{Re}(\sigma) - \eta - 
      2 \lambda)\Big) (-4  \mathbf{Re}(\sigma) + \eta + 2 \lambda)
\end{align}
First note again that there is always the trivial Floquet exponent $\eta=0$. For the second factor we find
\begin{equation}
\lambda_{P(AP)} =\kappa\frac{\mathbf{Re}(\gamma) \sin^{2}\phi+ 2 \mathbf{Im}(\gamma) \sin \phi \cos \phi + 3 \mathbf{Re}(\gamma)  \cos^2 \phi}{\mathbf{Im}(\gamma) \sin \phi  + \mathbf{Re}(\gamma) \cos \phi}
\end{equation} 
fulfills $\eta = 0$, corresponding to the anti-phase pitchfork of limit cycles ($P_{AP}$ in Fig.~\ref{2d_plot_multi}). Looking for purely imaginary $\eta$ we find
\begin{equation}
\lambda_{T(AP)} = 4 \mathbf{Re}(s) = 4 \kappa \cos\phi.
\end{equation} 
which describes the secondary Hopf bifurcation on the anti-phase solution ($T_{AP}$ in Fig.~\ref{2d_plot_multi}). The last factor simply recovers the generating Hopf-bifurcation of the anti-phase oscillations ($H_{AP}$ in Fig.~\ref{2d_plot_multi}) with $\lambda = 2 \kappa \cos(\phi)$.

These are all the bifurcations found for the off-state $Z_1 = Z_2 = 0$ and for the in-phase and anti-phase solutions. However, the Saddle-Node of limit cycles shown in Fig.~\ref{asymmmetric_states_fig} can not be derived this way, but emerges as a necessary condition when calculating the symmetry-broken amplitude- and phase-coupling states (see next section).

\subsection{Analytical symmetry-broken solutions}
\label{sec:app:analyt}

In the following we show the derivation of the analytical formulas for describing the symmetry-broken phase and amplitude locking states. In the general case we have two coupled ODEs:
\begin{align}\label{slsupp}
\dot{Z}_{1}&= \left(\lambda+\ii \omega + \gamma\,|Z_{1}|^{2}\right) Z_{1}+\sigma \left(Z_{2}-Z_{1}\right)\\
\dot{Z}_{2}&= \left(\lambda+\ii \omega + \gamma\,|Z_{2}|^{2}\right) Z_{2}+\sigma \left(Z_{1}-Z_{2}\right). \label{slsupp2}
\end{align}
We redefine $\Lambda = \lambda - \mathbf{Re}(\sigma)$ and $\tilde{\omega} = \omega -\mathbf{Im}(\sigma)$. We are interested in symmetry-broken states $
Z_{2}= s Z_{1} := s Z$
with constant $s \in \mathbb{C}$. We are only interested in states of pure harmonic oscillations, i.e., rotations in the complex plain. We can therefore make the rotating wave ansatz $\dot{Z}= \ii \tilde{\Omega} Z$
which leads to
\begin{align}
0 &= \left(\Lambda+\ii \Omega + \gamma\,|Z|^{2}\right) + s \sigma \label{Supp_Eq_1}\\
0 &= \left(\Lambda+\ii \Omega + \gamma\,|s|^{2} |Z|^{2}\right) s  + \sigma \label{Supp_Eq_2},
\end{align}
where $ \Omega = \tilde{\omega} - \tilde{\Omega}$. We have also assumed $Z\not = 0$ and divided by $Z$. We have four real-valued unknowns left: $|Z|^2$, $\Omega$, $\mathbf{Re}(s)$ and $\mathbf{Im}(s)$, for which the complex Eqs.~\eqref{Supp_Eq_1} and~\eqref{Supp_Eq_2} suffice. One can also transform these by taking Eq.~\eqref{Supp_Eq_1} multiplied by $s$ and subtracting Eq.~\eqref{Supp_Eq_2} and vice versa, leading to:
\begin{align}
0 &= \left(1 - |s|^{2} \right) s \, \gamma\,|Z|^{2} + \left(s^2 - 1\right) \sigma  \label{Supp_Eq_3}\\
0 &= \left(s^2 - 1\right) \left(\Lambda+\ii \Omega\right)  + \left( |s|^{2} s ^2 - 1 \right) \gamma\, |Z|^{2} s \label{Supp_Eq_4}
\end{align}
Equations~\eqref{Supp_Eq_3} and~\eqref{Supp_Eq_4} nicely illustrate that $s = \pm 1$ is always a solution of the system, corresponding to the in-phase and anti-phase orbits of coupled Stuart-Landau oscillators. Note, that we can not make use of the Fundamental Theorem of Algebra to reduce this equation, even though these roots are known, as Eqs.~\eqref{Supp_Eq_3} and~\eqref{Supp_Eq_4} contain non-polynomial elements $|s|^2$. 

We can use the real and imaginary part for Eq.~\eqref{Supp_Eq_1} to determine $|Z|^2$ and $\Omega$:
\begin{align}
|Z|^{2} &= \frac{-\Lambda - \mathbf{Re}(\sigma s) }{\mathbf{Re}(\gamma)}  \\
\Omega &= c \left(-\Lambda - \mathbf{Re}(\sigma s) \right) + \mathbf{Im}(\sigma s),
\end{align}
with $c = \mathbf{Im}(\gamma) /\mathbf{Re}(\gamma) $. After inserting these equations into Eq.~\eqref{Supp_Eq_4} we once again split into real and imaginary parts. One may be tempted to use the polar description of $s = a \exp (\ii \psi)$ to solve these equations, and indeed one can obtain Eq.~\eqref{a_of_psi_equation} linking $a$ and $\psi$ by this ansatz. However, we deem it impossible to fully solve for $a$ and $\psi$ this way, because of the transcendent nature of equations that are obtained. We therefore use $ s = \mathbf{Re}(s) + \ii \mathbf{Im}(s)$. Similarly, we set $\sigma = \alpha + \ii \beta$. This way, closed polynomial equations for $\mathbf{Re}(s)$ and $\mathbf{Im}(s)$ can be deduced. With additional definitions $x = \mathbf{Re}(s)$, $y = \mathbf{Im}(s)$, $g = \alpha / \beta$ and $L = \Lambda/\beta$ for visual clarity, these two equations for $s$ result in the following equations for $x$ and $y$:
\begin{align}
0 =& \left( g \, c - 1 \right) \left( x^3 + x\, y^2 - x \right) - \left(g + c\right) \left( y^3 + x^2 \, y + y\right)  \label{Supp_Eq_5} \\
0 =& \left( - L - g \, x + y  \right) \left( y + c \, x  \right) \left( 1 - x^2 - y^2 \right) \nonumber \\
& + 2 g\, x\,y + x^2 - y^2  - 1, \label{Supp_Eq_6}
\end{align}
where we have used the definitions  Eqs.~\eqref{Supp_Eq_5} and~\eqref{Supp_Eq_6} lead to a final quartic equation that are difficult to solve by hand. We have used the \textit{Solve}-function of \textit{Wolfram Mathematica} to solve Eqs.~\eqref{Supp_Eq_5} and~\eqref{Supp_Eq_6}. Apart from $s = \pm 1$, we also obtain solutions of the form:
\begin{align}
\mathbf{Re}(s)_{1, 2} =  - T_1 - \frac{1}{2}\sqrt{T_2} \pm \frac{1}{2}\sqrt{T_3} \label{Supp_Eq_7}\\
\mathbf{Re}(s)_{3, 4} =  - T_1 + \frac{1}{2}\sqrt{T_2} \pm \frac{1}{2}\sqrt{T_4}, \label{Supp_Eq_8}
\end{align}
with the following definitions:
\begin{align}
T_1 &= \frac{L (c+g) \left(c^2+4 c g-3\right)}{4 \left(g^2+1\right) \left(c^2+2 c g-1\right)} \\
T_2 &= \frac{(c+g)^2 \left(\left(c^2+1\right)^2 L^2-8 (c g-1) \left(c^2+2 c g-1\right)\right)}{4 \left(g^2+1\right)^2 \left(c^2+2 c g-1\right)^2} \\
T_{3, 4} &= \frac{(c+g)}{2 \left(g^2+1\right)^2 \left(c^2+2 c g-1\right)^2}  \nonumber \\
& \times \Big[-2 \left(g^2+1\right) (c+g) \left(c^2 \left(L^2+2\right)+4 c g+L^2-2\right)  \nonumber \\
& \pm \left(g^2+1\right) L \left(c^2+2 c g-1\right) \left[(c+2 g)^2+1\right] \sqrt{4 F_2}    \nonumber \\
& + L^2 \left(c^2+4 c g-3\right)^2 (c+g) -4 g \left(c^2+2 c g-1\right) (c+g)^2  \nonumber \\ 
& -2 L^2 \left(2 c^3 g+c^2 \left(5 g^2-1\right)-6 c g+g^2+3\right) (c+g)\Big]
\end{align}
We still need to be able to produce a similarly clear description for $\mathbf{Im}(s)$, but two ways of obtaining it from $\mathbf{Re}(s)$ are possible: First, we can use the $a$-to-$\psi$ formula obtained earlier in the manuscript and rewrite it to obtain $\mathbf{Im}(s)$ as a function of $\mathbf{Re}(s)$. Alternatively, we can use the fact that for each $s$ the corresponding value $1/s$ is also a solution to our system of equations. Hence $s_1 = 1/s_2$ and $s_3 = 1/s_4$. Consequently, we do know $\mathbf{Re}(s)$ and $\mathbf{Re}(1/s)$ and through some simple algebra can obtain $\mathbf{Im}(s)$. All of these approaches introduce some spurious solutions one needs to be careful not to include, for which we have mainly relied on numerical simulations in the paper. 

Additionally, by splitting $s$ into real and imaginary part, we know that $\mathbf{Re}(s)$ and $\mathbf{Im}(s)$ need to be real-valued. Hence we can use the square roots inside Eqs.~\eqref{Supp_Eq_7} and~\eqref{Supp_Eq_8} to obtain necessary conditions for the existence of symmetry-broken amplitude and phase locking states. By solving $T_2 \geq 0$ we obtain the bifurcation line of the saddle-node of limit cycles (Eq.~\eqref{SN}) as described in the paper and plotted as the red line in Fig.~\ref{asymmmetric_states_fig}.

\subsection{Special cases $\psi=0,\pi$ and $\psi=\pi/2,\, 3 \pi/2$}
\label{sec:app:special_cases}

We are describing symmetry-broken amplitude and phase-locking states $Z_1 = a e^{\ii \psi} Z_2$ with $a \not = 1$ or $\psi \not =0, \, \pi$. First, we will search for solutions where the oscillators exhibit harmonic waves with different radius but with phase shifts restricted to $\psi=0$ or $\psi=\pi$, i.e., we search for nontrivial in- and anti-phase solutions fulfilling  $Z_{1} = \pm a  Z_{2}$ which we plug into Eq.~\eqref{slsupp} and~\eqref{slsupp2}.
We simplify and obtain the following two equations for $Z_{2}$:
\begin{align} \label{blb}
\dot{Z}_{2}&= \left(\lambda+\ii \omega  + \gamma\, a^{2}\,|Z_{2}|^{2}\right)   Z_{2} +\sigma \left(\pm Z_{2} /a- Z_{2} \right)\\
\dot{Z}_{2}&= \left(\lambda+\ii \omega  + \gamma\,|Z_{2}|^{2}\right) Z_{2}+\sigma \left(\pm a \, Z_{2} -Z_{2}\right). \label{blb_part2}
\end{align}
The oscillator $Z_{2}$ needs to fulfill both equations in order to give us a solution of the type we are looking for.
In principle, each equation is a shifted Hopf normal form, so we need to find out where the branches of periodic solutions for each equation meet. 
In the following, we omit the index of $Z_{2}$ and only write $Z$ for simplicity.
Let us investigate Eq.~\eqref{blb} first which we rearrange to find:
\begin{align}
\dot{Z}= &  \left(\lambda \pm\mathbf{Re}(\sigma) /a- \mathbf{Re}(\sigma)\right) Z \nonumber \\ 
  + & \ii \left(\omega \pm \mathbf{Im}(\sigma) /a-  \mathbf{Im}(\sigma)  \right)Z+\gamma\, a^{2}\,|Z|^{2} Z   
\end{align}
We denote 
\begin{align}
\Lambda_{1}=&\lambda \pm\mathbf{Re}(\sigma) /a- \mathbf{Re}(\sigma), \\
\Omega_{1}=&\omega \pm \mathbf{Im}(\sigma) /a-  \mathbf{Im}(\sigma), \\
\Gamma_{1}=&\gamma \, a^{2}.
\end{align}
Then the bifurcating periodic orbit is given by its radius $R_1$ and its frequency $F_1$:
\begin{equation}
R_{1}=\sqrt{-\Lambda_{1}/\Re\Gamma_{1}},\qquad
F_{1}=\Omega_{1} -\Lambda_{1}  \mathbf{Im}(\Gamma)_{1}/\mathbf{Re}(\Gamma)_{1}.
\end{equation}
We proceed analogously for the second equation~\eqref{blb_part2}:
\begin{align}
\dot{Z}=& \left(\lambda \pm a \mathbf{Re}(\sigma) - \mathbf{Re}(\sigma)\right) Z \nonumber \\
 &+\ii \left(\omega \pm a \Im \sigma-  \mathbf{Im}(\sigma)  \right)Z+\gamma\, |Z|^{2} Z   
\end{align}
Here we denote 
\begin{align}
\Lambda_{2}=&\lambda \pm a \mathbf{Re}(\sigma) - \mathbf{Re}(\sigma), \nonumber \\
\Omega_{2}=&\omega\pm a \mathbf{Im}(\sigma) -  \mathbf{Im}(\sigma), \nonumber \\
\Gamma_{2}=&\gamma 
\end{align}
In this case, the bifurcating periodic orbit is given by its radius $R_2$ and its frequency $F_2$:
\begin{equation}
R_{2}=\sqrt{-\Lambda_{2}/\mathbf{Re}(\Gamma) _{2}},\qquad
F_{2}=\Omega_{2} -\Lambda_{2}  \mathbf{Im}(\Gamma)_{2}/\mathbf{Re}(\Gamma) _{2}.
\end{equation}
The task is to find a parameter $a$ such that $R_{1}=R_{2}$ and $F_{1}=F_{2}$.
Let us start with $R_{1}=R_{2}$, i.e., $R_{1}^{2}=R_{2}^{2}$. We obtain the following fourth order polynomial in $a$:
\begin{equation}\label{fourth}
0=\mp\, a^{4} \mathbf{Re}(\sigma) +  a^{3}(\mathbf{Re}(\sigma)-\lambda)   +a (\lambda  - \mathbf{Re}(\sigma))\pm \mathbf{Re}(\sigma) 
\end{equation}
Next, we solve for $F_{1}=F_{2}$ where we obtain a quadratic equation in $a$:
\begin{equation}
0= (a^{2}-1) \left(-\Im \sigma  + \mathbf{Re}(\sigma)  \frac{ \mathbf{Im}(\gamma)}{\mathbf{Re}(\gamma) }\right)
\end{equation}
This gives us either the condition $1=a^{2}$,
which is consistent with the known in-phase and anti-phase solutions of the system or the condition
\begin{equation}
\frac{ \mathbf{Im}(\gamma)}{\mathbf{Re}(\gamma) }= \frac{\mathbf{Im}(\sigma)}{\mathbf{Re}(\sigma)},
\end{equation}
as claimed in the paper.
We insert this into Eq. \eqref{fourth} to obtain the branches $a(\lambda)$ of Eq.~\eqref{special_case_branches_1} and Eq.~\eqref{special_case_branches_2}. These are the branches for nontrivial in-phase and anti-phase solutions.

In the same way we will now search for solutions with phase shifts restricted to $\psi= \pi/2$ or $\psi=3\pi/2$, i.e., we search for nontrivial solutions fulfilling  $Z_{1} = \ii a  Z_{2}$ which we plug into Eq.~\eqref{slsupp}.
We simplify and obtain the following two equations for $Z_{2}$:
\begin{align} \label{blb2}
\dot{Z}_{2}&= \left(\lambda+\ii \omega  + \gamma\, a^{2}\,|Z_{2}|^{2}\right)   Z_{2} +\sigma \left(-\ii Z_{2} /a- Z_{2} \right)\\
\dot{Z}_{2}&= \left(\lambda+\ii \omega  + \gamma\,|Z_{2}|^{2}\right) Z_{2}+\sigma \left( \ii a  \, Z_{2} -Z_{2}\right). \label{blb2-part2}
\end{align}
Again, the oscillator $Z_{2}$ needs to fulfill both equations in order to give us a solution of the type we are looking for.
For simplicity, we omit the index of $Z_{2}$ and only write $Z$.
We carry out the same calculations as above and find the bifurcating periodic orbit, given by its radius $\tilde{R}_1$ and its frequency $\tilde{F}_1$: 
\begin{align}
\tilde{R}_{1}  =& \sqrt{-\left(\lambda +\mathbf{Re}(\sigma) /a- \mathbf{Im}(\sigma)\right)/\left(\gamma \, a^{2}\right)} \\
\tilde{F}_{1}  =& \omega - \mathbf{Re}(\sigma) /a-  \mathbf{Im}(\sigma) \nonumber \\ &-\left(\lambda +\mathbf{Re}(\sigma) /a- \mathbf{Im}(\sigma)\right) \mathbf{Im}(\gamma)/\mathbf{Re}(\gamma).
\end{align}
We proceed completely analogously for the second equation~\eqref{blb2-part2}:
\begin{align}
\tilde{R}_{2}=&\sqrt{-\left(\lambda +a\, \mathbf{Im}(\sigma) + \mathbf{Re}(\sigma)\right)/\left(\gamma \, a^{2}\right)} \\\tilde{F}_{2}=&\omega + a \mathbf{Re}(\sigma) -  \mathbf{Im}(\sigma) \nonumber \\ &-\left(\lambda -a\, \mathbf{Im}(\sigma) - \mathbf{Re}(\sigma)\right) \mathbf{Im}(\gamma)/\mathbf{Re}(\gamma).\end{align}
We set $\tilde{R}_{1}=\tilde{R}_{2}$, and obtain the following fourth order polynomial in $a$:
\begin{equation}\label{fourth2}
0= a^{4} \mathbf{Im}(\sigma) +  a^{3}(\mathbf{Re}(\sigma)-\lambda)   +a (\lambda  - \mathbf{Re}(\sigma))+ \mathbf{Im}(\sigma) 
\end{equation}
Next, we set $\tilde{F}_{1}=\tilde{F}_{2}$ which yields a quadratic equation in $a$:
\begin{equation}
0= (a^{2}+1) \left( \mathbf{Re} \sigma  + \mathbf{Im}(\sigma)  \frac{ \mathbf{Im}(\gamma)}{\mathbf{Re}(\gamma) }\right)
\end{equation}
This gives us the condition $-1=a^{2}$, which is again consistent with the known in-phase and anti-phase solutions  of the system, as this would lead to $Z_1 = \ii a Z_2 = \pm Z_2$.  Alternatively the condition
\begin{equation}
\frac{ \mathbf{Im}(\gamma)}{\mathbf{Re}(\gamma) }= -\frac{\mathbf{Re}(\sigma)}{\mathbf{Im}(\sigma)},
\end{equation}
arises as claimed in the paper.
For this specific set of parameters, we find two distinct branches of solutions that exist for the asymmetric states $Z_{1} = \ii a  Z_{2}$ given by 
\begin{align}
& a(\lambda) =
-\frac{1}{2} \sqrt{-2 + \frac{(\mathbf{Re}(\sigma) - \lambda)^2}{4 \,\mathbf{Im}(\sigma)^2}} - 
\frac{ \mathbf{Re}(\sigma) - \lambda}{4\,\mathbf{Im}(\sigma)} \nonumber \\
 &\pm\frac{1}{2} \sqrt{2 + \frac{(\mathbf{Re}(\sigma)- \lambda)^2}{
   2\, \mathbf{Im}(\sigma)^2} + \frac{\frac{(\mathbf{Re}(\sigma) - \lambda)^3}{\mathbf{Im}(\sigma)^2} + 8
(\mathbf{Re}(\sigma) + \lambda)}{
    \sqrt{-32 \mathbf{Im}(\sigma)^2 + 4(\mathbf{Re}(\sigma) - \lambda)^2}}},\\
&    a(\lambda) =\hphantom{-}
\frac{1}{2} \sqrt{-2 + \frac{(\mathbf{Re}(\sigma) - \lambda)^2}{4 \,\mathbf{Im}(\sigma)^2}} - 
\frac{ \mathbf{Re}(\sigma) - \lambda}{4\,\mathbf{Im}(\sigma)} \nonumber \\
 &\pm\frac{1}{2} \sqrt{2 + \frac{(\mathbf{Re}(\sigma)- \lambda)^2}{
   2\, \mathbf{Im}(\sigma)^2} + \frac{\frac{(\mathbf{Re}(\sigma) - \lambda)^3}{\mathbf{Im}(\sigma)^2} + 8
(\mathbf{Re}(\sigma) + \lambda)}{
    \sqrt{-32 \mathbf{Im}(\sigma)^2 + 4(\mathbf{Re}(\sigma) - \lambda)^2}}}.
\end{align}
Note that these branches do not cross any pitchfork bifurcation, as the coupling phase values $\phi$ where these special $\psi = \pi/2,\, 3\pi/2$ solutions exist lead to the pitchfork bifurcation being pushed to $\lambda \rightarrow \infty$. The branches only emerge out of saddle-node bifurcations, see the grey dotted lines in Fig-~\ref{asymmmetric_states_fig}.

\subsection{Comparison with Globally Coupled Stuart-Landau Networks}
\label{model_compare}

Finally, we will shortly show the connection of the discussed solutions and parameter regions with cluster solutions in large networks. In previous publications of large globally coupled networks of Stuart-Landau oscillators, the authors often employ a different parametrization of the system \cite{HAK92}:

\begin{equation}
	\dot{A_j} = \frac{1 + \ii \beta}{N} \sum\limits_{k = 1}^{N} \left( A_k - A_j\right)  + \mu A_j  -  \left( 1 + \ii \alpha \right) |A_j|^2 A_j,   
	\label{Hakim_Rappelt_Equation}
\end{equation}

for the more general case of $N$ oscillators coupled all-to-all. This reduces to Eq.~\eqref{two_stuart_landau_equation_1} and~\eqref{two_stuart_landau_equation_2} for $N=2$ with the following substitutions: $\kappa \cos \phi = (1 + \ii \beta )/2$, $\lambda = \mu$, $\gamma = - 1 - \ii \alpha$, $\omega = 0$. In this representation, the frequency has been taken out by a corresponding rotating frame and the real part of the coupling has been normalized to 1 without loss of generality. In other parts of the literature the parameter $\lambda$ has been set to $1$ instead \cite{KU15}. 

\begin{figure}
        \includegraphics[width=1.0\linewidth]{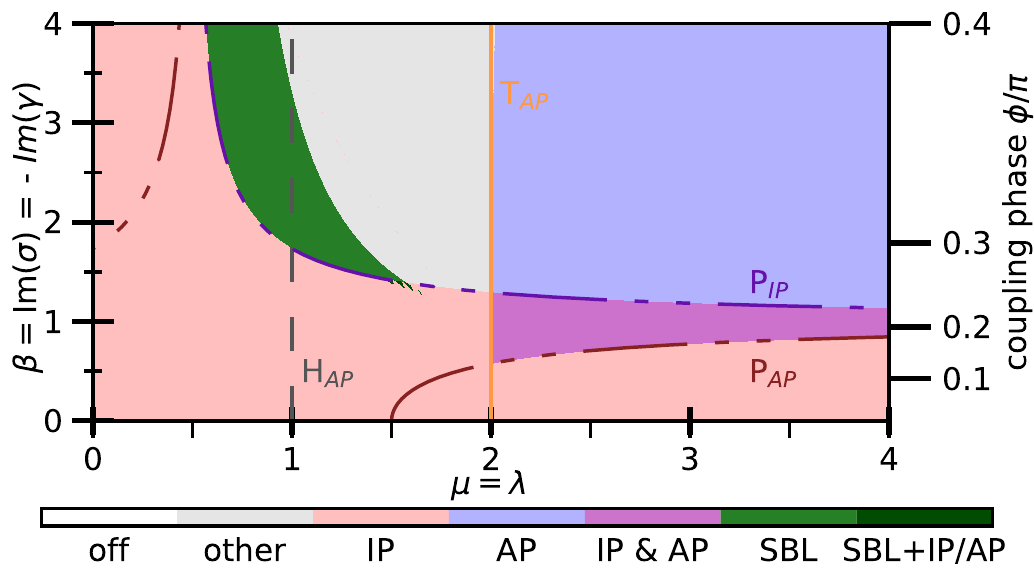}
        \caption{Numerically evaluated states of two coupled Stuart-Landau oscillators described by parameters according to Eq.~\eqref{Hakim_Rappelt_Equation}, to be compared with Fig.~1 from Ref.~\cite{HAK92}. bifurcation lines are our analytic results described by Eq.~\eqref{HAP} for the Hopf ($H_{AP}$), Eq.~\eqref{PIP} and~\eqref{PAP} for the pitchfork ($P_{IP}$ and $P_{AP}$), and Eq.~\eqref{TAP} for the torus bifurcation ($T_{AP}$). Parameters: $\alpha = - \beta$}
        \label{HAK_figure}
\end{figure}

To illustrate the relationship to our results, we have used the same evaluation methods from the main paper and plotted the resulting solutions for the system as described in Eq.~\eqref{Hakim_Rappelt_Equation} for $N=2$ and $\alpha = -\beta$. As can be seen in Fig.~\ref{HAK_figure}, we find the same boundaries for the synchronized and anti-synchronized regions as described in Ref.~\cite{HAK92}. However, due to the choice of parameters, the anti-phase synchronized solution never undergoes a supercritical Hopf-bifurcation in this representation.

The regions outside of these two main solutions have been investigated in great detail in Ref.~\cite{HAK92, NAK93, NAK94, KU15}, with a focus on chaotic motion and cluster formation. We find the symmetry-broken amplitude- and phase-locking states in the corresponding regions for $N=2$ here, suggesting an intimate connection. However, as far as we can tell, for a system with large $N$ no stable clusters of identical size but different amplitudes have actually been observed or predicted. This hints at possible internal instability of clusters on the corresponding branches of the symmetry-broken amplitude- and phase-locked states, or that the dynamics are no longer harmonic (also cf. Fig.~4 in Ref.~\cite{KU15} for low $f_a$). Cluster solutions of different sizes, i.e. different numbers of oscillators, are possibly not directly related to the solutions investigated in this manuscript. The connection is likely more complex than that, as these cluster states in large networks are more easily explained by a non-symmetric version of Eq.~\eqref{sl_derivation_ref1} and~\eqref{two_stuart_landau_equation_2} with weights, as pointed out by Ref.~\cite{HAK92}.

\section*{References}


\begin{thebibliography}{55}%
\makeatletter
\providecommand \@ifxundefined [1]{%
 \@ifx{#1\undefined}
}%
\providecommand \@ifnum [1]{%
 \ifnum #1\expandafter \@firstoftwo
 \else \expandafter \@secondoftwo
 \fi
}%
\providecommand \@ifx [1]{%
 \ifx #1\expandafter \@firstoftwo
 \else \expandafter \@secondoftwo
 \fi
}%
\providecommand \natexlab [1]{#1}%
\providecommand \enquote  [1]{``#1''}%
\providecommand \bibnamefont  [1]{#1}%
\providecommand \bibfnamefont [1]{#1}%
\providecommand \citenamefont [1]{#1}%
\providecommand \href@noop [0]{\@secondoftwo}%
\providecommand \href [0]{\begingroup \@sanitize@url \@href}%
\providecommand \@href[1]{\@@startlink{#1}\@@href}%
\providecommand \@@href[1]{\endgroup#1\@@endlink}%
\providecommand \@sanitize@url [0]{\catcode `\\12\catcode `\$12\catcode
  `\&12\catcode `\#12\catcode `\^12\catcode `\_12\catcode `\%12\relax}%
\providecommand \@@startlink[1]{}%
\providecommand \@@endlink[0]{}%
\providecommand \url  [0]{\begingroup\@sanitize@url \@url }%
\providecommand \@url [1]{\endgroup\@href {#1}{\urlprefix }}%
\providecommand \urlprefix  [0]{URL }%
\providecommand \Eprint [0]{\href }%
\providecommand \doibase [0]{http://dx.doi.org/}%
\providecommand \selectlanguage [0]{\@gobble}%
\providecommand \bibinfo  [0]{\@secondoftwo}%
\providecommand \bibfield  [0]{\@secondoftwo}%
\providecommand \translation [1]{[#1]}%
\providecommand \BibitemOpen [0]{}%
\providecommand \bibitemStop [0]{}%
\providecommand \bibitemNoStop [0]{.\EOS\space}%
\providecommand \EOS [0]{\spacefactor3000\relax}%
\providecommand \BibitemShut  [1]{\csname bibitem#1\endcsname}%
\let\auto@bib@innerbib\@empty
\bibitem [{\citenamefont {Lotka}(1920)}]{LOT20}%
  \BibitemOpen
  \bibfield  {author} {\bibinfo {author} {\bibfnamefont {A.~J.}\ \bibnamefont
  {Lotka}},\ }\href@noop {} {\bibfield  {journal} {\bibinfo  {journal} {Proc.
  Natl. Acad. Sci. U.S.A.}\ }\textbf {\bibinfo {volume} {6}} (\bibinfo {year}
  {1920})}\BibitemShut {NoStop}%
\bibitem [{\citenamefont {Volterra}(1926)}]{VOL26a}%
  \BibitemOpen
  \bibfield  {author} {\bibinfo {author} {\bibfnamefont {V.}~\bibnamefont
  {Volterra}},\ }\href {\doibase 10.1038/118558a0} {\bibfield  {journal}
  {\bibinfo  {journal} {Nature}\ }\textbf {\bibinfo {volume} {118}},\ \bibinfo
  {pages} {558} (\bibinfo {year} {1926})}\BibitemShut {NoStop}%
\bibitem [{\citenamefont {Belousov}(1959)}]{BEL59}%
  \BibitemOpen
  \bibfield  {author} {\bibinfo {author} {\bibfnamefont {B.~P.}\ \bibnamefont
  {Belousov}},\ }in\ \href@noop {} {\emph {\bibinfo {booktitle} {Collection of
  short papers on radiation medicine for 1958}}}\ (\bibinfo  {publisher} {Med.
  Publ.},\ \bibinfo {address} {Moscow},\ \bibinfo {year} {1959})\BibitemShut
  {NoStop}%
\bibitem [{\citenamefont {Zhabotinskii}(1964)}]{ZHA64}%
  \BibitemOpen
  \bibfield  {author} {\bibinfo {author} {\bibfnamefont {A.~M.}\ \bibnamefont
  {Zhabotinskii}},\ }\href@noop {} {\bibfield  {journal} {\bibinfo  {journal}
  {Biofizika}\ }\textbf {\bibinfo {volume} {9}},\ \bibinfo {pages} {306}
  (\bibinfo {year} {1964})}\BibitemShut {NoStop}%
\bibitem [{\citenamefont {Erneux}\ and\ \citenamefont
  {Glorieux}(2010)}]{ERN10b}%
  \BibitemOpen
  \bibfield  {author} {\bibinfo {author} {\bibfnamefont {T.}~\bibnamefont
  {Erneux}}\ and\ \bibinfo {author} {\bibfnamefont {P.}~\bibnamefont
  {Glorieux}},\ }\href@noop {} {\emph {\bibinfo {title} {Laser Dynamics}}}\
  (\bibinfo  {publisher} {Cambridge University Press},\ \bibinfo {address}
  {UK},\ \bibinfo {year} {2010})\BibitemShut {NoStop}%
\bibitem [{\citenamefont {Garcia-Morales}\ and\ \citenamefont
  {Krischer}(2012)}]{GAR12b}%
  \BibitemOpen
  \bibfield  {author} {\bibinfo {author} {\bibfnamefont {V.}~\bibnamefont
  {Garcia-Morales}}\ and\ \bibinfo {author} {\bibfnamefont {K.}~\bibnamefont
  {Krischer}},\ }\href@noop {} {\bibfield  {journal} {\bibinfo  {journal}
  {Contemporary Physics}\ }\textbf {\bibinfo {volume} {53}},\ \bibinfo {pages}
  {79} (\bibinfo {year} {2012})}\BibitemShut {NoStop}%
\bibitem [{\citenamefont {Hopf}(1942)}]{HOP42}%
  \BibitemOpen
  \bibfield  {author} {\bibinfo {author} {\bibfnamefont {E.}~\bibnamefont
  {Hopf}},\ }\href@noop {} {\bibfield  {journal} {\bibinfo  {journal}
  {Ber.Math.-Phys. Klasse Sachs. Akad. Wiss.}\ }\textbf {\bibinfo {volume}
  {94}},\ \bibinfo {pages} {1} (\bibinfo {year} {1942})}\BibitemShut {NoStop}%
\bibitem [{\citenamefont {Andronov}\ and\ \citenamefont
  {Chajkin}(1949)}]{AND49}%
  \BibitemOpen
  \bibfield  {author} {\bibinfo {author} {\bibfnamefont {A.~A.}\ \bibnamefont
  {Andronov}}\ and\ \bibinfo {author} {\bibfnamefont {S.~E.}\ \bibnamefont
  {Chajkin}},\ }\href@noop {} {\emph {\bibinfo {title} {Theory of
  oscillations.}}},\ edited by\ \bibinfo {editor} {\bibfnamefont
  {S.}~\bibnamefont {Lefschetz}}\ (\bibinfo  {publisher} {Princeton Univ.
  Press},\ \bibinfo {year} {1949})\BibitemShut {NoStop}%
\bibitem [{\citenamefont {Kuramoto}(1984)}]{KUR84}%
  \BibitemOpen
  \bibfield  {author} {\bibinfo {author} {\bibfnamefont {Y.}~\bibnamefont
  {Kuramoto}},\ }\href@noop {} {\emph {\bibinfo {title} {Chemical Oscillations,
  Waves and Turbulence}}}\ (\bibinfo  {publisher} {Springer-Verlag},\ \bibinfo
  {address} {Berlin},\ \bibinfo {year} {1984})\BibitemShut {NoStop}%
\bibitem [{\citenamefont {Kuznetsov}(1995)}]{KUZ95}%
  \BibitemOpen
  \bibfield  {author} {\bibinfo {author} {\bibfnamefont {Y.~A.}\ \bibnamefont
  {Kuznetsov}},\ }\href@noop {} {\emph {\bibinfo {title} {Elements of Applied
  Bifurcation Theory}}}\ (\bibinfo  {publisher} {Springer},\ \bibinfo {address}
  {New York},\ \bibinfo {year} {1995})\BibitemShut {NoStop}%
\bibitem [{\citenamefont {Soriano}\ \emph {et~al.}(2013)\citenamefont
  {Soriano}, \citenamefont {Garc{\'i}a-Ojalvo}, \citenamefont {Mirasso},\ and\
  \citenamefont {Fischer}}]{SOR13}%
  \BibitemOpen
  \bibfield  {author} {\bibinfo {author} {\bibfnamefont {M.~C.}\ \bibnamefont
  {Soriano}}, \bibinfo {author} {\bibfnamefont {J.}~\bibnamefont
  {Garc{\'i}a-Ojalvo}}, \bibinfo {author} {\bibfnamefont {C.~R.}\ \bibnamefont
  {Mirasso}}, \ and\ \bibinfo {author} {\bibfnamefont {I.}~\bibnamefont
  {Fischer}},\ }\href@noop {} {\bibfield  {journal} {\bibinfo  {journal}
  {Rev.~Mod.~Phys.}\ }\textbf {\bibinfo {volume} {85}},\ \bibinfo {pages} {421}
  (\bibinfo {year} {2013})}\BibitemShut {NoStop}%
\bibitem [{\citenamefont {Garc\'\i{}a-Ojalvo}\ \emph
  {et~al.}(1999)\citenamefont {Garc\'\i{}a-Ojalvo}, \citenamefont {Casademunt},
  \citenamefont {Torrent}, \citenamefont {Mirasso},\ and\ \citenamefont
  {Sancho}}]{GAR99a}%
  \BibitemOpen
  \bibfield  {author} {\bibinfo {author} {\bibfnamefont {J.}~\bibnamefont
  {Garc\'\i{}a-Ojalvo}}, \bibinfo {author} {\bibfnamefont {J.}~\bibnamefont
  {Casademunt}}, \bibinfo {author} {\bibfnamefont {M.~C.}\ \bibnamefont
  {Torrent}}, \bibinfo {author} {\bibfnamefont {C.~R.}\ \bibnamefont
  {Mirasso}}, \ and\ \bibinfo {author} {\bibfnamefont {J.~M.}\ \bibnamefont
  {Sancho}},\ }\href {\doibase 10.1142/s021812749900167x} {\bibfield  {journal}
  {\bibinfo  {journal} {Int. J. Bifurcation Chaos}\ }\textbf {\bibinfo {volume}
  {9}},\ \bibinfo {pages} {2225} (\bibinfo {year} {1999})}\BibitemShut
  {NoStop}%
\bibitem [{\citenamefont {Javaloyes}\ \emph {et~al.}(2003)\citenamefont
  {Javaloyes}, \citenamefont {Mandel},\ and\ \citenamefont {Pieroux}}]{JAV03}%
  \BibitemOpen
  \bibfield  {author} {\bibinfo {author} {\bibfnamefont {J.}~\bibnamefont
  {Javaloyes}}, \bibinfo {author} {\bibfnamefont {P.}~\bibnamefont {Mandel}}, \
  and\ \bibinfo {author} {\bibfnamefont {D.}~\bibnamefont {Pieroux}},\
  }\href@noop {} {\bibfield  {journal} {\bibinfo  {journal} {Phys. Rev. E}\
  }\textbf {\bibinfo {volume} {67}},\ \bibinfo {pages} {036201} (\bibinfo
  {year} {2003})}\BibitemShut {NoStop}%
\bibitem [{\citenamefont {Yanchuk}\ \emph {et~al.}(2004)\citenamefont
  {Yanchuk}, \citenamefont {Schneider},\ and\ \citenamefont {Recke}}]{YAN04c}%
  \BibitemOpen
  \bibfield  {author} {\bibinfo {author} {\bibfnamefont {S.}~\bibnamefont
  {Yanchuk}}, \bibinfo {author} {\bibfnamefont {K.~R.}\ \bibnamefont
  {Schneider}}, \ and\ \bibinfo {author} {\bibfnamefont {L.}~\bibnamefont
  {Recke}},\ }\href {\doibase 10.1103/physreve.69.056221} {\bibfield  {journal}
  {\bibinfo  {journal} {Phys. Rev. E}\ }\textbf {\bibinfo {volume} {69}},\
  \bibinfo {pages} {056221} (\bibinfo {year} {2004})}\BibitemShut {NoStop}%
\bibitem [{\citenamefont {Clerkin}\ \emph {et~al.}(2014)\citenamefont
  {Clerkin}, \citenamefont {O'Brien},\ and\ \citenamefont {Amann}}]{CLE14}%
  \BibitemOpen
  \bibfield  {author} {\bibinfo {author} {\bibfnamefont {E.}~\bibnamefont
  {Clerkin}}, \bibinfo {author} {\bibfnamefont {S.}~\bibnamefont {O'Brien}}, \
  and\ \bibinfo {author} {\bibfnamefont {A.}~\bibnamefont {Amann}},\
  }\href@noop {} {\bibfield  {journal} {\bibinfo  {journal} {Phys. Rev. E}\
  }\textbf {\bibinfo {volume} {89}},\ \bibinfo {pages} {032919} (\bibinfo
  {year} {2014})}\BibitemShut {NoStop}%
\bibitem [{\citenamefont {Kozyreff}\ \emph {et~al.}(2000)\citenamefont
  {Kozyreff}, \citenamefont {Vladimirov},\ and\ \citenamefont
  {Mandel}}]{KOZ00}%
  \BibitemOpen
  \bibfield  {author} {\bibinfo {author} {\bibfnamefont {G.}~\bibnamefont
  {Kozyreff}}, \bibinfo {author} {\bibfnamefont {A.~G.}\ \bibnamefont
  {Vladimirov}}, \ and\ \bibinfo {author} {\bibfnamefont {P.}~\bibnamefont
  {Mandel}},\ }\href@noop {} {\bibfield  {journal} {\bibinfo  {journal}
  {Phys.~Rev.~Lett.}\ }\textbf {\bibinfo {volume} {85}},\ \bibinfo {pages}
  {3809} (\bibinfo {year} {2000})}\BibitemShut {NoStop}%
\bibitem [{\citenamefont {Erzgr{\"a}ber}\ \emph {et~al.}(2005)\citenamefont
  {Erzgr{\"a}ber}, \citenamefont {Lenstra}, \citenamefont {Krauskopf},
  \citenamefont {Wille}, \citenamefont {Peil}, \citenamefont {Fischer},\ and\
  \citenamefont {Els{\"a}{\ss}er}}]{ERZ05}%
  \BibitemOpen
  \bibfield  {author} {\bibinfo {author} {\bibfnamefont {H.}~\bibnamefont
  {Erzgr{\"a}ber}}, \bibinfo {author} {\bibfnamefont {D.}~\bibnamefont
  {Lenstra}}, \bibinfo {author} {\bibfnamefont {B.}~\bibnamefont {Krauskopf}},
  \bibinfo {author} {\bibfnamefont {E.}~\bibnamefont {Wille}}, \bibinfo
  {author} {\bibfnamefont {M.}~\bibnamefont {Peil}}, \bibinfo {author}
  {\bibfnamefont {I.}~\bibnamefont {Fischer}}, \ and\ \bibinfo {author}
  {\bibfnamefont {W.}~\bibnamefont {Els{\"a}{\ss}er}},\ }\href@noop {}
  {\bibfield  {journal} {\bibinfo  {journal} {Opt. Commun.}\ }\textbf {\bibinfo
  {volume} {255}},\ \bibinfo {pages} {286} (\bibinfo {year}
  {2005})}\BibitemShut {NoStop}%
\bibitem [{\citenamefont {Erzgr{\"a}ber}\ \emph {et~al.}(2009)\citenamefont
  {Erzgr{\"a}ber}, \citenamefont {Wille}, \citenamefont {Krauskopf},\ and\
  \citenamefont {Fischer}}]{ERZ09}%
  \BibitemOpen
  \bibfield  {author} {\bibinfo {author} {\bibfnamefont {H.}~\bibnamefont
  {Erzgr{\"a}ber}}, \bibinfo {author} {\bibfnamefont {E.}~\bibnamefont
  {Wille}}, \bibinfo {author} {\bibfnamefont {B.}~\bibnamefont {Krauskopf}}, \
  and\ \bibinfo {author} {\bibfnamefont {I.}~\bibnamefont {Fischer}},\
  }\href@noop {} {\bibfield  {journal} {\bibinfo  {journal} {Nonlinearity}\
  }\textbf {\bibinfo {volume} {22}},\ \bibinfo {pages} {585} (\bibinfo {year}
  {2009})}\BibitemShut {NoStop}%
\bibitem [{\citenamefont {Hohl}\ \emph {et~al.}(1999)\citenamefont {Hohl},
  \citenamefont {Gavrielides}, \citenamefont {Erneux},\ and\ \citenamefont
  {Kovanis}}]{HOH99b}%
  \BibitemOpen
  \bibfield  {author} {\bibinfo {author} {\bibfnamefont {A.}~\bibnamefont
  {Hohl}}, \bibinfo {author} {\bibfnamefont {A.}~\bibnamefont {Gavrielides}},
  \bibinfo {author} {\bibfnamefont {T.}~\bibnamefont {Erneux}}, \ and\ \bibinfo
  {author} {\bibfnamefont {V.}~\bibnamefont {Kovanis}},\ }\href {\doibase
  10.1103/physreva.59.3941} {\bibfield  {journal} {\bibinfo  {journal}
  {Phys.~Rev.~A}\ }\textbf {\bibinfo {volume} {59}},\ \bibinfo {pages} {3941}
  (\bibinfo {year} {1999})}\BibitemShut {NoStop}%
\bibitem [{\citenamefont {Shim}\ \emph {et~al.}(2007)\citenamefont {Shim},
  \citenamefont {Imboden},\ and\ \citenamefont {Mohanty}}]{SHI07a}%
  \BibitemOpen
  \bibfield  {author} {\bibinfo {author} {\bibfnamefont {S.~B.}\ \bibnamefont
  {Shim}}, \bibinfo {author} {\bibfnamefont {M.}~\bibnamefont {Imboden}}, \
  and\ \bibinfo {author} {\bibfnamefont {P.}~\bibnamefont {Mohanty}},\ }\href
  {\doibase 10.1126/science.1137307} {\bibfield  {journal} {\bibinfo  {journal}
  {Science}\ }\textbf {\bibinfo {volume} {316}},\ \bibinfo {pages} {95}
  (\bibinfo {year} {2007})}\BibitemShut {NoStop}%
\bibitem [{\citenamefont {Totz}\ \emph {et~al.}(2015)\citenamefont {Totz},
  \citenamefont {Snari}, \citenamefont {Yengi}, \citenamefont {Tinsley},
  \citenamefont {Engel},\ and\ \citenamefont {Showalter}}]{TOT15}%
  \BibitemOpen
  \bibfield  {author} {\bibinfo {author} {\bibfnamefont {J.}~\bibnamefont
  {Totz}}, \bibinfo {author} {\bibfnamefont {R.}~\bibnamefont {Snari}},
  \bibinfo {author} {\bibfnamefont {D.}~\bibnamefont {Yengi}}, \bibinfo
  {author} {\bibfnamefont {M.}~\bibnamefont {Tinsley}}, \bibinfo {author}
  {\bibfnamefont {H.}~\bibnamefont {Engel}}, \ and\ \bibinfo {author}
  {\bibfnamefont {K.}~\bibnamefont {Showalter}},\ }\href {\doibase
  10.1103/physreve.92.022819} {\bibfield  {journal} {\bibinfo  {journal} {Phys.
  Rev. E}\ }\textbf {\bibinfo {volume} {92}},\ \bibinfo {pages} {022819}
  (\bibinfo {year} {2015})}\BibitemShut {NoStop}%
\bibitem [{\citenamefont {Weiner}\ \emph {et~al.}(1992)\citenamefont {Weiner},
  \citenamefont {Holz}, \citenamefont {Schneider},\ and\ \citenamefont
  {Bar-Eli}}]{WEI92}%
  \BibitemOpen
  \bibfield  {author} {\bibinfo {author} {\bibfnamefont {J.}~\bibnamefont
  {Weiner}}, \bibinfo {author} {\bibfnamefont {R.}~\bibnamefont {Holz}},
  \bibinfo {author} {\bibfnamefont {F.~W.}\ \bibnamefont {Schneider}}, \ and\
  \bibinfo {author} {\bibfnamefont {K.}~\bibnamefont {Bar-Eli}},\ }\href
  {\doibase 10.1021/j100201a041} {\bibfield  {journal} {\bibinfo  {journal}
  {J.~Phys.~Chem.}\ }\textbf {\bibinfo {volume} {96}},\ \bibinfo {pages} {8915}
  (\bibinfo {year} {1992})}\BibitemShut {NoStop}%
\bibitem [{\citenamefont {Heinrich}\ \emph {et~al.}(2010)\citenamefont
  {Heinrich}, \citenamefont {Dahms}, \citenamefont {Flunkert}, \citenamefont
  {Teitsworth},\ and\ \citenamefont {Sch{\"o}ll}}]{HEI10}%
  \BibitemOpen
  \bibfield  {author} {\bibinfo {author} {\bibfnamefont {M.}~\bibnamefont
  {Heinrich}}, \bibinfo {author} {\bibfnamefont {T.}~\bibnamefont {Dahms}},
  \bibinfo {author} {\bibfnamefont {V.}~\bibnamefont {Flunkert}}, \bibinfo
  {author} {\bibfnamefont {S.~W.}\ \bibnamefont {Teitsworth}}, \ and\ \bibinfo
  {author} {\bibfnamefont {E.}~\bibnamefont {Sch{\"o}ll}},\ }\href {\doibase
  10.1088/1367-2630/12/11/113030} {\bibfield  {journal} {\bibinfo  {journal}
  {New~J.~Phys.}\ }\textbf {\bibinfo {volume} {12}},\ \bibinfo {pages} {113030}
  (\bibinfo {year} {2010})}\BibitemShut {NoStop}%
\bibitem [{\citenamefont {Kapitaniak}\ \emph {et~al.}(2012)\citenamefont
  {Kapitaniak}, \citenamefont {Czolczynski}, \citenamefont {Perlikowski},
  \citenamefont {Stefanski},\ and\ \citenamefont {Kapitaniak}}]{KAP12}%
  \BibitemOpen
  \bibfield  {author} {\bibinfo {author} {\bibfnamefont {M.}~\bibnamefont
  {Kapitaniak}}, \bibinfo {author} {\bibfnamefont {K.}~\bibnamefont
  {Czolczynski}}, \bibinfo {author} {\bibfnamefont {P.}~\bibnamefont
  {Perlikowski}}, \bibinfo {author} {\bibfnamefont {A.}~\bibnamefont
  {Stefanski}}, \ and\ \bibinfo {author} {\bibfnamefont {T.}~\bibnamefont
  {Kapitaniak}},\ }\href@noop {} {\bibfield  {journal} {\bibinfo  {journal}
  {Phys. Rep.}\ }\textbf {\bibinfo {volume} {517}},\ \bibinfo {pages} {1}
  (\bibinfo {year} {2012})}\BibitemShut {NoStop}%
\bibitem [{\citenamefont {Kuramoto}\ and\ \citenamefont
  {Battogtokh}(2002)}]{KUR02a}%
  \BibitemOpen
  \bibfield  {author} {\bibinfo {author} {\bibfnamefont {Y.}~\bibnamefont
  {Kuramoto}}\ and\ \bibinfo {author} {\bibfnamefont {D.}~\bibnamefont
  {Battogtokh}},\ }\href@noop {} {\bibfield  {journal} {\bibinfo  {journal}
  {Nonlin. Phen. in Complex Sys.}\ }\textbf {\bibinfo {volume} {5}},\ \bibinfo
  {pages} {380} (\bibinfo {year} {2002})}\BibitemShut {NoStop}%
\bibitem [{\citenamefont {Abrams}\ and\ \citenamefont
  {Strogatz}(2004)}]{ABR04}%
  \BibitemOpen
  \bibfield  {author} {\bibinfo {author} {\bibfnamefont {D.~M.}\ \bibnamefont
  {Abrams}}\ and\ \bibinfo {author} {\bibfnamefont {S.~H.}\ \bibnamefont
  {Strogatz}},\ }\href {\doibase 10.1103/physrevlett.93.174102} {\bibfield
  {journal} {\bibinfo  {journal} {Phys.~Rev.~Lett.}\ }\textbf {\bibinfo
  {volume} {93}},\ \bibinfo {pages} {174102} (\bibinfo {year}
  {2004})}\BibitemShut {NoStop}%
\bibitem [{\citenamefont {{D'Huys}}\ \emph {et~al.}(2008)\citenamefont
  {{D'Huys}}, \citenamefont {Vicente}, \citenamefont {Erneux}, \citenamefont
  {Danckaert},\ and\ \citenamefont {Fischer}}]{DHU08}%
  \BibitemOpen
  \bibfield  {author} {\bibinfo {author} {\bibfnamefont {O.}~\bibnamefont
  {{D'Huys}}}, \bibinfo {author} {\bibfnamefont {R.}~\bibnamefont {Vicente}},
  \bibinfo {author} {\bibfnamefont {T.}~\bibnamefont {Erneux}}, \bibinfo
  {author} {\bibfnamefont {J.}~\bibnamefont {Danckaert}}, \ and\ \bibinfo
  {author} {\bibfnamefont {I.}~\bibnamefont {Fischer}},\ }\href {\doibase
  10.1063/1.2953582} {\bibfield  {journal} {\bibinfo  {journal} {Chaos}\
  }\textbf {\bibinfo {volume} {18}},\ \bibinfo {pages} {037116} (\bibinfo
  {year} {2008})}\BibitemShut {NoStop}%
\bibitem [{\citenamefont {Kaneko}(1990)}]{KAN90}%
  \BibitemOpen
  \bibfield  {author} {\bibinfo {author} {\bibfnamefont {K.}~\bibnamefont
  {Kaneko}},\ }\href {\doibase doi: 10.1016/0167-2789(90)90119-a} {\bibfield
  {journal} {\bibinfo  {journal} {Physica D}\ }\textbf {\bibinfo {volume}
  {41}},\ \bibinfo {pages} {137} (\bibinfo {year} {1990})}\BibitemShut
  {NoStop}%
\bibitem [{\citenamefont {Hakim}\ and\ \citenamefont {Rappel}(1992)}]{HAK92}%
  \BibitemOpen
  \bibfield  {author} {\bibinfo {author} {\bibfnamefont {V.}~\bibnamefont
  {Hakim}}\ and\ \bibinfo {author} {\bibfnamefont {W.~J.}\ \bibnamefont
  {Rappel}},\ }\href {\doibase 10.1103/physreva.46.r7347} {\bibfield  {journal}
  {\bibinfo  {journal} {Phys. Rev. A}\ }\textbf {\bibinfo {volume} {46}},\
  \bibinfo {pages} {R7347} (\bibinfo {year} {1992})}\BibitemShut {NoStop}%
\bibitem [{\citenamefont {Nakagawa}\ and\ \citenamefont
  {Kuramoto}(1993)}]{NAK93}%
  \BibitemOpen
  \bibfield  {author} {\bibinfo {author} {\bibfnamefont {K.}~\bibnamefont
  {Nakagawa}}\ and\ \bibinfo {author} {\bibfnamefont {Y.}~\bibnamefont
  {Kuramoto}},\ }\href@noop {} {\bibfield  {journal} {\bibinfo  {journal}
  {Prog. Theor. Phys.}\ }\textbf {\bibinfo {volume} {89}},\ \bibinfo {pages}
  {313} (\bibinfo {year} {1993})}\BibitemShut {NoStop}%
\bibitem [{\citenamefont {Nakagawa}\ and\ \citenamefont
  {Kuramoto}(1994)}]{NAK94a}%
  \BibitemOpen
  \bibfield  {author} {\bibinfo {author} {\bibfnamefont {N.}~\bibnamefont
  {Nakagawa}}\ and\ \bibinfo {author} {\bibfnamefont {Y.}~\bibnamefont
  {Kuramoto}},\ }\href@noop {} {\bibfield  {journal} {\bibinfo  {journal}
  {Physica D}\ }\textbf {\bibinfo {volume} {75}},\ \bibinfo {pages} {74}
  (\bibinfo {year} {1994})}\BibitemShut {NoStop}%
\bibitem [{\citenamefont {Ku}\ \emph {et~al.}(2015)\citenamefont {Ku},
  \citenamefont {Girvan},\ and\ \citenamefont {Ott}}]{KU15}%
  \BibitemOpen
  \bibfield  {author} {\bibinfo {author} {\bibfnamefont {W.~L.}\ \bibnamefont
  {Ku}}, \bibinfo {author} {\bibfnamefont {M.}~\bibnamefont {Girvan}}, \ and\
  \bibinfo {author} {\bibfnamefont {E.}~\bibnamefont {Ott}},\ }\href@noop {}
  {\bibfield  {journal} {\bibinfo  {journal} {Chaos}\ }\textbf {\bibinfo
  {volume} {25}} (\bibinfo {year} {2015})}\BibitemShut {NoStop}%
\bibitem [{\citenamefont {Koseska}\ \emph {et~al.}(2013)\citenamefont
  {Koseska}, \citenamefont {Volkov},\ and\ \citenamefont {Kurths}}]{KOS13}%
  \BibitemOpen
  \bibfield  {author} {\bibinfo {author} {\bibfnamefont {A.}~\bibnamefont
  {Koseska}}, \bibinfo {author} {\bibfnamefont {E.}~\bibnamefont {Volkov}}, \
  and\ \bibinfo {author} {\bibfnamefont {J.}~\bibnamefont {Kurths}},\
  }\href@noop {} {\bibfield  {journal} {\bibinfo  {journal} {Phys. Rep.}\
  }\textbf {\bibinfo {volume} {531}},\ \bibinfo {pages} {173} (\bibinfo {year}
  {2013})}\BibitemShut {NoStop}%
\bibitem [{\citenamefont {Zakharova}\ \emph {et~al.}(2013)\citenamefont
  {Zakharova}, \citenamefont {Feoktistov}, \citenamefont {Vadivasova},\ and\
  \citenamefont {Sch{\"o}ll}}]{ZAK13}%
  \BibitemOpen
  \bibfield  {author} {\bibinfo {author} {\bibfnamefont {A.}~\bibnamefont
  {Zakharova}}, \bibinfo {author} {\bibfnamefont {A.}~\bibnamefont
  {Feoktistov}}, \bibinfo {author} {\bibfnamefont {T.}~\bibnamefont
  {Vadivasova}}, \ and\ \bibinfo {author} {\bibfnamefont {E.}~\bibnamefont
  {Sch{\"o}ll}},\ }\href {\doibase 10.1140/epjst/e2013-02031-x} {\bibfield
  {journal} {\bibinfo  {journal} {Eur. Phys. J. Spec. Top.}\ }\textbf {\bibinfo
  {volume} {222}},\ \bibinfo {pages} {2481} (\bibinfo {year}
  {2013})}\BibitemShut {NoStop}%
\bibitem [{\citenamefont {Ushakov}\ \emph {et~al.}(2005)\citenamefont
  {Ushakov}, \citenamefont {W{\"u}nsche}, \citenamefont {Henneberger},
  \citenamefont {Khovanov}, \citenamefont {Schimansky-Geier},\ and\
  \citenamefont {Zaks}}]{USH05}%
  \BibitemOpen
  \bibfield  {author} {\bibinfo {author} {\bibfnamefont {O.~V.}\ \bibnamefont
  {Ushakov}}, \bibinfo {author} {\bibfnamefont {H.~J.}\ \bibnamefont
  {W{\"u}nsche}}, \bibinfo {author} {\bibfnamefont {F.}~\bibnamefont
  {Henneberger}}, \bibinfo {author} {\bibfnamefont {I.~A.}\ \bibnamefont
  {Khovanov}}, \bibinfo {author} {\bibfnamefont {L.}~\bibnamefont
  {Schimansky-Geier}}, \ and\ \bibinfo {author} {\bibfnamefont {M.~A.}\
  \bibnamefont {Zaks}},\ }\href@noop {} {\bibfield  {journal} {\bibinfo
  {journal} {Phys.~Rev.~Lett.}\ }\textbf {\bibinfo {volume} {95}},\ \bibinfo
  {pages} {123903} (\bibinfo {year} {2005})}\BibitemShut {NoStop}%
\bibitem [{\citenamefont {Geffert}\ \emph {et~al.}(2014)\citenamefont
  {Geffert}, \citenamefont {Zakharova}, \citenamefont {V{\"u}llings},
  \citenamefont {Just},\ and\ \citenamefont {Sch{\"o}ll}}]{GEF14}%
  \BibitemOpen
  \bibfield  {author} {\bibinfo {author} {\bibfnamefont {P.~M.}\ \bibnamefont
  {Geffert}}, \bibinfo {author} {\bibfnamefont {A.}~\bibnamefont {Zakharova}},
  \bibinfo {author} {\bibfnamefont {A.}~\bibnamefont {V{\"u}llings}}, \bibinfo
  {author} {\bibfnamefont {W.}~\bibnamefont {Just}}, \ and\ \bibinfo {author}
  {\bibfnamefont {E.}~\bibnamefont {Sch{\"o}ll}},\ }\href {\doibase
  10.1140/epjb/e2014-50541-2} {\bibfield  {journal} {\bibinfo  {journal} {Eur.
  Phys. J.~B}\ }\textbf {\bibinfo {volume} {87}},\ \bibinfo {pages} {291}
  (\bibinfo {year} {2014})}\BibitemShut {NoStop}%
\bibitem [{\citenamefont {Zakharova}\ \emph {et~al.}(2014)\citenamefont
  {Zakharova}, \citenamefont {Kapeller},\ and\ \citenamefont
  {Sch{\"o}ll}}]{ZAK14}%
  \BibitemOpen
  \bibfield  {author} {\bibinfo {author} {\bibfnamefont {A.}~\bibnamefont
  {Zakharova}}, \bibinfo {author} {\bibfnamefont {M.}~\bibnamefont {Kapeller}},
  \ and\ \bibinfo {author} {\bibfnamefont {E.}~\bibnamefont {Sch{\"o}ll}},\
  }\href {\doibase 10.1103/physrevlett.112.154101} {\bibfield  {journal}
  {\bibinfo  {journal} {Phys.~Rev.~Lett.}\ }\textbf {\bibinfo {volume} {112}},\
  \bibinfo {pages} {154101} (\bibinfo {year} {2014})}\BibitemShut {NoStop}%
\bibitem [{\citenamefont {Zakharova}\ \emph {et~al.}(2016)\citenamefont
  {Zakharova}, \citenamefont {Kapeller},\ and\ \citenamefont
  {Sch{\"o}ll}}]{ZAK15b}%
  \BibitemOpen
  \bibfield  {author} {\bibinfo {author} {\bibfnamefont {A.}~\bibnamefont
  {Zakharova}}, \bibinfo {author} {\bibfnamefont {M.}~\bibnamefont {Kapeller}},
  \ and\ \bibinfo {author} {\bibfnamefont {E.}~\bibnamefont {Sch{\"o}ll}},\
  }\href@noop {} {\bibfield  {journal} {\bibinfo  {journal} {J. Phys. Conf.
  Series}\ }\textbf {\bibinfo {volume} {727}},\ \bibinfo {pages} {012018}
  (\bibinfo {year} {2016})}\BibitemShut {NoStop}%
\bibitem [{\citenamefont {Lee}\ \emph {et~al.}(2013)\citenamefont {Lee},
  \citenamefont {Ott},\ and\ \citenamefont {Antonsen}}]{LEE13}%
  \BibitemOpen
  \bibfield  {author} {\bibinfo {author} {\bibfnamefont {W.~S.}\ \bibnamefont
  {Lee}}, \bibinfo {author} {\bibfnamefont {E.}~\bibnamefont {Ott}}, \ and\
  \bibinfo {author} {\bibfnamefont {T.~M.}\ \bibnamefont {Antonsen}},\
  }\href@noop {} {\bibfield  {journal} {\bibinfo  {journal} {Chaos}\ }\textbf
  {\bibinfo {volume} {23}},\ \bibinfo {pages} {033116} (\bibinfo {year}
  {2013})}\BibitemShut {NoStop}%
\bibitem [{\citenamefont {Lehnert}\ \emph {et~al.}(2014)\citenamefont
  {Lehnert}, \citenamefont {H{\"o}vel}, \citenamefont {Selivanov},
  \citenamefont {Fradkov},\ and\ \citenamefont {Sch{\"o}ll}}]{LEH14}%
  \BibitemOpen
  \bibfield  {author} {\bibinfo {author} {\bibfnamefont {J.}~\bibnamefont
  {Lehnert}}, \bibinfo {author} {\bibfnamefont {P.}~\bibnamefont {H{\"o}vel}},
  \bibinfo {author} {\bibfnamefont {A.~A.}\ \bibnamefont {Selivanov}}, \bibinfo
  {author} {\bibfnamefont {A.~L.}\ \bibnamefont {Fradkov}}, \ and\ \bibinfo
  {author} {\bibfnamefont {E.}~\bibnamefont {Sch{\"o}ll}},\ }\href {\doibase
  10.1103/physreve.90.042914} {\bibfield  {journal} {\bibinfo  {journal}
  {Phys.~Rev.~E}\ }\textbf {\bibinfo {volume} {90}},\ \bibinfo {pages} {042914}
  (\bibinfo {year} {2014})}\BibitemShut {NoStop}%
\bibitem [{\citenamefont {Golubitsky}\ and\ \citenamefont
  {Stewart}(1988)}]{GOL88a}%
  \BibitemOpen
  \bibfield  {author} {\bibinfo {author} {\bibfnamefont {M.}~\bibnamefont
  {Golubitsky}}\ and\ \bibinfo {author} {\bibfnamefont {I.}~\bibnamefont
  {Stewart}},\ }\href@noop {} {\emph {\bibinfo {title} {Singularities and
  Groups in Bifurcation Theory. Volume 2}}},\ Vol.~\bibinfo {volume} {69}\
  (\bibinfo  {publisher} {Springer-Verlag},\ \bibinfo {address} {New York},\
  \bibinfo {year} {1988})\BibitemShut {NoStop}%
\bibitem [{\citenamefont {Seifikar}\ \emph {et~al.}(2017)\citenamefont
  {Seifikar}, \citenamefont {Amann},\ and\ \citenamefont {Peters}}]{SEI17}%
  \BibitemOpen
  \bibfield  {author} {\bibinfo {author} {\bibfnamefont {M.}~\bibnamefont
  {Seifikar}}, \bibinfo {author} {\bibfnamefont {A.}~\bibnamefont {Amann}}, \
  and\ \bibinfo {author} {\bibfnamefont {F.~H.}\ \bibnamefont {Peters}},\
  }\href {\doibase 10.1051/ epjconf/201713900010} {\bibfield  {journal}
  {\bibinfo  {journal} {EPJ Web of Conferences}\ }\textbf {\bibinfo {volume}
  {139}} (\bibinfo {year} {2017}),\ 10.1051/ epjconf/201713900010}\BibitemShut
  {NoStop}%
\bibitem [{\citenamefont {B{\"o}hm}\ \emph {et~al.}(2015)\citenamefont
  {B{\"o}hm}, \citenamefont {Zakharova}, \citenamefont {Sch{\"o}ll},\ and\
  \citenamefont {L{\"u}dge}}]{BOE15}%
  \BibitemOpen
  \bibfield  {author} {\bibinfo {author} {\bibfnamefont {F.}~\bibnamefont
  {B{\"o}hm}}, \bibinfo {author} {\bibfnamefont {A.}~\bibnamefont {Zakharova}},
  \bibinfo {author} {\bibfnamefont {E.}~\bibnamefont {Sch{\"o}ll}}, \ and\
  \bibinfo {author} {\bibfnamefont {K.}~\bibnamefont {L{\"u}dge}},\ }\href@noop
  {} {\bibfield  {journal} {\bibinfo  {journal} {Phys. Rev. E}\ }\textbf
  {\bibinfo {volume} {91}},\ \bibinfo {pages} {040901 (R)} (\bibinfo {year}
  {2015})}\BibitemShut {NoStop}%
\bibitem [{\citenamefont {Aronson}\ \emph {et~al.}(1990)\citenamefont
  {Aronson}, \citenamefont {Ermentrout},\ and\ \citenamefont {Kopell}}]{ARO90}%
  \BibitemOpen
  \bibfield  {author} {\bibinfo {author} {\bibfnamefont {D.~G.}\ \bibnamefont
  {Aronson}}, \bibinfo {author} {\bibfnamefont {G.~B.}\ \bibnamefont
  {Ermentrout}}, \ and\ \bibinfo {author} {\bibfnamefont {N.}~\bibnamefont
  {Kopell}},\ }\href@noop {} {\bibfield  {journal} {\bibinfo  {journal}
  {Physica~D}\ }\textbf {\bibinfo {volume} {41}},\ \bibinfo {pages} {403}
  (\bibinfo {year} {1990})}\BibitemShut {NoStop}%
\bibitem [{\citenamefont {B{\"o}hm}\ and\ \citenamefont
  {L{\"u}dge}(2016)}]{BOE16}%
  \BibitemOpen
  \bibfield  {author} {\bibinfo {author} {\bibfnamefont {F.}~\bibnamefont
  {B{\"o}hm}}\ and\ \bibinfo {author} {\bibfnamefont {K.}~\bibnamefont
  {L{\"u}dge}},\ }in\ \href@noop {} {\emph {\bibinfo {booktitle} {Control of
  Self-Organizing Nonlinear Systems}}},\ Vol.\ \bibinfo {volume} {Chapter 18}\
  (\bibinfo  {publisher} {Springer},\ \bibinfo {year} {2016})\ Chap.~\bibinfo
  {chapter} {18}, pp.\ \bibinfo {pages} {1--20}\BibitemShut {NoStop}%
\bibitem [{\citenamefont {R{\"o}hm}\ \emph {et~al.}(2016)\citenamefont
  {R{\"o}hm}, \citenamefont {B{\"o}hm},\ and\ \citenamefont
  {L{\"u}dge}}]{ROE16}%
  \BibitemOpen
  \bibfield  {author} {\bibinfo {author} {\bibfnamefont {A.}~\bibnamefont
  {R{\"o}hm}}, \bibinfo {author} {\bibfnamefont {F.}~\bibnamefont {B{\"o}hm}},
  \ and\ \bibinfo {author} {\bibfnamefont {K.}~\bibnamefont {L{\"u}dge}},\
  }\href {\doibase 10.1103/physreve.94.042204} {\bibfield  {journal} {\bibinfo
  {journal} {Phys. Rev. E}\ }\textbf {\bibinfo {volume} {94}},\ \bibinfo
  {pages} {042204} (\bibinfo {year} {2016})}\BibitemShut {NoStop}%
\bibitem [{\citenamefont {Spencer}\ and\ \citenamefont {Lamb}(1972)}]{SPE72}%
  \BibitemOpen
  \bibfield  {author} {\bibinfo {author} {\bibfnamefont {M.~B.}\ \bibnamefont
  {Spencer}}\ and\ \bibinfo {author} {\bibfnamefont {W.~E.}\ \bibnamefont
  {Lamb}, \bibfnamefont {Jr.}},\ }\href {\doibase 10.1103/physreva.5.893}
  {\bibfield  {journal} {\bibinfo  {journal} {Phys. Rev.~A}\ }\textbf {\bibinfo
  {volume} {5}},\ \bibinfo {pages} {893} (\bibinfo {year} {1972})}\BibitemShut
  {NoStop}%
\bibitem [{\citenamefont {Pikovsky}\ \emph {et~al.}(2001)\citenamefont
  {Pikovsky}, \citenamefont {Rosenblum},\ and\ \citenamefont {Kurths}}]{PIK01}%
  \BibitemOpen
  \bibfield  {author} {\bibinfo {author} {\bibfnamefont {A.}~\bibnamefont
  {Pikovsky}}, \bibinfo {author} {\bibfnamefont {M.~G.}\ \bibnamefont
  {Rosenblum}}, \ and\ \bibinfo {author} {\bibfnamefont {J.}~\bibnamefont
  {Kurths}},\ }\href@noop {} {\emph {\bibinfo {title} {Synchronization, A
  Universal Concept in Nonlinear Sciences}}}\ (\bibinfo  {publisher} {Cambridge
  University Press},\ \bibinfo {address} {Cambridge},\ \bibinfo {year}
  {2001})\BibitemShut {NoStop}%
\bibitem [{\citenamefont {Crawford}(1991)}]{CRA91}%
  \BibitemOpen
  \bibfield  {author} {\bibinfo {author} {\bibfnamefont {J.~D.}\ \bibnamefont
  {Crawford}},\ }\href@noop {} {\bibfield  {journal} {\bibinfo  {journal} {Rev.
  Mod. Phys.}\ }\textbf {\bibinfo {volume} {63}},\ \bibinfo {pages} {991}
  (\bibinfo {year} {1991})}\BibitemShut {NoStop}%
\bibitem [{\citenamefont {Hoyle}(2006)}]{HOY06}%
  \BibitemOpen
  \bibfield  {author} {\bibinfo {author} {\bibfnamefont {R.}~\bibnamefont
  {Hoyle}},\ }\href@noop {} {\emph {\bibinfo {title} {Pattern Formation: An
  Introduction to Methods}}},\ Cambridge Texts in Applied Mathematics\
  (\bibinfo  {publisher} {Cambridge University Press},\ \bibinfo {year}
  {2006})\BibitemShut {NoStop}%
\bibitem [{\citenamefont {Schmidt}\ and\ \citenamefont
  {Krischer}(2014)}]{SCH14n}%
  \BibitemOpen
  \bibfield  {author} {\bibinfo {author} {\bibfnamefont {L.}~\bibnamefont
  {Schmidt}}\ and\ \bibinfo {author} {\bibfnamefont {K.}~\bibnamefont
  {Krischer}},\ }\href {\doibase 10.1103/physreve.90.042911} {\bibfield
  {journal} {\bibinfo  {journal} {Phys. Rev.~E}\ }\textbf {\bibinfo {volume}
  {90}} (\bibinfo {year} {2014}),\ 10.1103/physreve.90.042911}\BibitemShut
  {NoStop}%
\bibitem [{\citenamefont {Kominis}\ \emph {et~al.}(2017)\citenamefont
  {Kominis}, \citenamefont {Kovanis},\ and\ \citenamefont {Bountis}}]{KOM17}%
  \BibitemOpen
  \bibfield  {author} {\bibinfo {author} {\bibfnamefont {Y.}~\bibnamefont
  {Kominis}}, \bibinfo {author} {\bibfnamefont {V.}~\bibnamefont {Kovanis}}, \
  and\ \bibinfo {author} {\bibfnamefont {T.}~\bibnamefont {Bountis}},\
  }\href@noop {} {\bibfield  {journal} {\bibinfo  {journal} {Phys. Rev.~A}\
  }\textbf {\bibinfo {volume} {96}},\ \bibinfo {pages} {043836} (\bibinfo
  {year} {2017})}\BibitemShut {NoStop}%
\bibitem [{\citenamefont {Lingnau}\ and\ \citenamefont
  {L{\"u}dge}(2015)}]{LIN15a}%
  \BibitemOpen
  \bibfield  {author} {\bibinfo {author} {\bibfnamefont {B.}~\bibnamefont
  {Lingnau}}\ and\ \bibinfo {author} {\bibfnamefont {K.}~\bibnamefont
  {L{\"u}dge}},\ }\href {\doibase 10.3390/photonics2020402} {\bibfield
  {journal} {\bibinfo  {journal} {Photonics}\ }\textbf {\bibinfo {volume}
  {2}},\ \bibinfo {pages} {402} (\bibinfo {year} {2015})}\BibitemShut {NoStop}%
\bibitem [{\citenamefont {Sivaramakrishnan}(2017)}]{SIV17}%
  \BibitemOpen
  \bibfield  {author} {\bibinfo {author} {\bibfnamefont {S.}~\bibnamefont
  {Sivaramakrishnan}},\ }\emph {\bibinfo {title} {Dynamics of Passively Coupled
  Continuous-Wave and Mode-Locked Lasers}},\ \href@noop {} {Ph.D. thesis},\
  \bibinfo  {school} {The University of Michigan} (\bibinfo {year}
  {2017})\BibitemShut {NoStop}%
\bibitem [{\citenamefont {Nakamura}\ \emph {et~al.}(1994)\citenamefont
  {Nakamura}, \citenamefont {Tominaga},\ and\ \citenamefont
  {Munakata}}]{NAK94}%
  \BibitemOpen
  \bibfield  {author} {\bibinfo {author} {\bibfnamefont {Y.}~\bibnamefont
  {Nakamura}}, \bibinfo {author} {\bibfnamefont {F.}~\bibnamefont {Tominaga}},
  \ and\ \bibinfo {author} {\bibfnamefont {T.}~\bibnamefont {Munakata}},\
  }\href {\doibase 10.1103/physreve.49.4849} {\bibfield  {journal} {\bibinfo
  {journal} {Phys. Rev. E}\ }\textbf {\bibinfo {volume} {49}},\ \bibinfo
  {pages} {4849} (\bibinfo {year} {1994})}\BibitemShut {NoStop}%
\end{thebibliography}

%

\end{document}